\documentclass[fleqn,usenatbib]{mnras}

\usepackage{graphicx}    % Including figure files
\usepackage{amsmath}    % Advanced maths commands
\usepackage{amssymb}    % Extra maths symbols
\usepackage{bm}% bold math
\usepackage{natbib}
\usepackage{subfigure}
\usepackage{color,units}
\usepackage{float}
\usepackage{aas_macros}
\usepackage{hyperref}
\usepackage[T1]{fontenc}
\usepackage{ae,aecompl}

\usepackage{totcount}
\newtotcounter{citnum} %From the package documentation
\def\oldbibitem{} \let\oldbibitem=\bibitem
\def\bibitem{\stepcounter{citnum}\oldbibitem}
\newcommand{\bilby}{{\sc Bilby}}
\newcommand{\dynesty}{{\sc dynesty}}
\newcommand{\afterglowpy}{{\sc afterglowpy}}
\newcommand{\cocoon}{{cocoon}}
\newcommand{\konus}{{KONUS-\textit{Wind}}}
\newcommand{\thetaobs}{{3^\circ}^{+3^\circ}_{-2^\circ}}
\newcommand{\thetacore}{{8^\circ} \pm 2^{\circ}}
\newcommand{\nism}{2^{+1}_{-2}}
\newcommand{\tstart}{58875.1^{+0.6}_{-1}}
\newcommand{\tp}{58876.1^{+0.1}_{-0.2}}
\newcommand{\eiso}{53.6^{+0.3}_{-0.4}}
\newcommand{\lorentz}{500\pm 300}

\newcommand{\efficiencyfermi}{0.3\%}
\newcommand{\efficiencykonus}{2.8\%}
\newcommand{\efficiencyipn}{4.5\%}
\newcommand{\percentagehundred}{7}
\newcommand{\percentagetwenty}{1}
\newcommand{\bfsplcocoon}{\sim 4}
\newcommand{\bfsploffaxis}{\sim 3\times10^{64}}
\newcommand{\bfgaussiancocoon}{\sim 4}
\newcommand{\bftophatcocoon}{\sim 6}
\newcommand{\splsigma}{7.7\times 10^{-4}}
\newcommand{\cocoonsigma}{8.2\times 10^{-4}}
\newcommand{\citeg}[1]{\citep[e.g.,][]{#1}}

\title[GRB origin of AT2020blt]{Low-efficiency long gamma-ray bursts: A case study with AT2020blt}

\author[N.Sarin et al.]{
\parbox{\textwidth}{
N.Sarin$^{1,2,3,4}$\thanks{E-mail:nikhil.sarin@su.se}, R.~Hamburg$^{5,6}$,E.~Burns$^{7}$, G.~Ashton$^{1,2,8}$, P.~D. Lasky$^{1,2}$ and G.~P.~Lamb$^{9}$} \vspace{0.2cm}\\
$^{1}$School of Physics and Astronomy, Monash University, Vic 3800, Australia\\
$^{2}$OzGrav: The A.R.C. Centre of Excellence for Gravitational Wave Discovery, Clayton VIC 3800, Australia\\
$^{3}$Nordita, KTH Royal Institute of Technology and Stockholm University Roslagstullsbacken 23, SE-106 91 Stockholm, Sweden\\
$^{4}$The Oskar Klein Centre, Department of Physics, Stockholm University, AlbaNova, SE-106 91 Stockholm, Sweden\\
$^{5}$Department of Space Science, University of Alabama in Huntsville, Huntsville, AL 35899, U.S.A. \\
$^{6}$Center for Space Plasma and Aeronomic Research, University of Alabama in Huntsville, Huntsville, AL 35899\\
$^{7}$Department of Physics and Astronomy, Louisiana State University, Baton Rouge, LA 70803, U.S.A. \\
$^{8}$Royal Holloway University of London, Egham, Surrey, TW20 0EX, U.K. \\
$^{9}$School of Physics and Astronomy, University of Leicester, University Road, LE1 7RH, U.K.}

\date{Accepted XXX. Received Y.Y.Y.; in original form ZZZ}
\pubyear{2021}

\begin{document}
\label{firstpage}
\pagerange{\pageref{firstpage}--\pageref{lastpage}}
\maketitle

\begin{abstract}
The Zwicky Transient Facility recently announced the detection of an optical transient AT2020blt at redshift $z=2.9$, consistent with the afterglow of an on-axis gamma-ray burst. However, no prompt emission was observed. We analyse AT2020blt with detailed models, showing the data are best explained as the afterglow of an on-axis long gamma-ray burst, ruling out other hypotheses such as a cocoon and a low-Lorentz factor jet. We search \textit{Fermi} data for prompt emission, setting deeper upper limits on the prompt emission than in the original detection paper. Together with KONUS-\textit{Wind} observations, we show that the gamma-ray efficiency of AT2020blt is $\lesssim 0.3 - 4.5\%$. We speculate that AT2020blt and AT2021any belong to the low-efficiency tail of long gamma-ray burst distributions that are beginning to be readily observed due to the capabilities of new observatories like the Zwicky Transient Facility. 
\end{abstract}

\begin{keywords}
gamma-ray bursts
\end{keywords}
%%%%%%%%%%%%%%%%% BODY OF PAPER %%%%%%%%%%%%%%%%%%
\section{Introduction}\label{sec:intro}
%No Bacon lettuce tomato sandwiches were consumed in the production of this manuscript.
The interaction of an ultra-relativistic jet launched in a gamma-ray burst with the surrounding interstellar medium is known to produce broadband synchrotron radiation referred to as an afterglow. Observed extensively in X rays, optical and radio, these phenomena are observationally confirmed to be linked to the collapse of massive stars and compact object mergers~\citep[e.g.,][]{cano17, abbott17_gw170817_gwgrb}. 

Afterglows of gamma-ray bursts are predominantly observed following-up the prompt emission trigger. There are a few exceptional cases such as iPTF11gg~\citep{cenko11}, iPTF14yb~\citep{cenko15}, FIRSTJ141918~\citep{marcote19}, and AT2020blt~\citep{ho20}. 
These transients were either independently detected from the prompt emission and later associated to a gamma-ray counterpart that were missed in low-latency~\citep{cenko15}, initially believed to be observed off-axis, in which case the high-energy prompt emission was missed due to relativistic beaming~\citep{Rhoads1997, Nakar2002, huang02}, or later shown to not be afterglows at all~\citep{lee20}.

The transient AT2020blt~\citep{ho20} is a fast-fading optical transient at $z=2.9$ observed by the Zwicky Transient Facility without any high-energy trigger from gamma-ray satellites. 
This transient has characteristic features akin to afterglows from gamma-ray bursts and broadband observations in X rays, optical, and radio~\citep{singer20,ho20}. 
However, the non-detection of prompt gamma-ray emission is puzzling.
\cite{ho20} suggest that this non-detection may be due to one of three reasons: 1) The prompt emission expected in gamma-rays was weak or missed due to occultation by the Earth. 2) The transient was the afterglow from an off-axis gamma-ray burst where the prompt gamma-rays were missed because of relativistic beaming. 
3) AT2020blt was the afterglow of a dirty fireball, i.e., a gamma-ray burst with a low Lorentz factor ($\Gamma \lesssim 100$), such that the optically-thick environment absorbed the gamma-ray photons.  

Prompt gamma-ray emission observed for both long and short gamma-ray bursts is likely produced by internal dissipation in an ultra-relativistic jet~\citep[e.g.,][]{kumar99}. 
% This may not have been true for GRB 170817A~\citep[e.g.,][]{ioka19,matsumoto19}, which may have produced off-axis prompt gamma-ray emission through a cocoon shock breakout~\citep[e.g.,][]{gottleib18}.
To produce a gamma-ray burst, the jet must be ultra-relativistic to avoided the compactness problem~\citep{ruderman75}, i.e., that gamma-ray photons are above the pair production threshold and should only be observable if the jet is moving ultra-relativistically. 
% To produce prompt gamma-ray emission, jets need to be ultra-relativistic.
% This constraint is necessitated by the gamma-ray burst compactness problem~\citep{ruderman75}, i.e., that gamma-ray photons are above the pair production threshold and should only be observable if the jet is moving ultra-relativistically. 
This theoretical constraint has led to the placement of lower limits on the initial bulk Lorentz factor of $\Gamma_0 \gtrsim 100$~\citeg{lithwick01}. The existence of relativistic jets following gamma-ray bursts has also been shown observationally, for e.g., through multi-wavelength observations of GRB 170817A~\citeg{Kasliwal2017,mooley18_superluminal, lamb18_gw170817, Ghirlanda2019, matsumoto19,Beniamini2020}.

As an ultra-relativistic jet passes through the stellar/ejecta envelope, it creates a \cocoon{} of shocked material. If the jet stalls within this envelope then such a \textit{choked} jet will dissipate energy into the surrounding bubble of matter, forming a quasi-spherical \textit{cocoon}, that will produce minimal prompt gamma-ray emission~\citeg{gottleib18}. 
% In general, the ultra-relativistic jet does not need to break out of the ejecta. Such a \textit{choked} jet will dissipate energy into the surrounding bubble of matter, forming a quasi-spherical \textit{cocoon}, that will produce minimal prompt gamma-ray emission~\citeg{gottleib18}. 
The interaction of the jet and/or \cocoon{} with the interstellar medium is ultimately responsible for producing the broadband afterglow we see following almost all gamma-ray bursts. 

The broadband observations of GRB 170817A confirmed that a relativistic jet successfully broke out of the ejecta and that the jet was likely structured i.e., the energy and Lorentz factor of the jet had some angular dependence. Although the exact jet structure is unknown, various phenomenological models such as a Gaussian or power-law structure can successfully explain the broadband afterglow observations of GRB 170817A~\citeg{troja17_xrays, lamb18_gw170817, Lazzati2018, Gill2018, Kathirgamaraju2018, afterglowpy}. We note that although a jet without any angular dependence is possible~\citep[e.g.,][]{aloy05}, it is unlikely as both the jet-launching mechanism and jet-breakout will likely produce some jet structure~\citep[e.g.,][]{nakar19}. 

In this paper, we investigate why there was no detection of prompt gamma-rays from AT2020blt by analysing the multi-wavelength data from~\cite{ho20} with detailed afterglow models. 
In Sec.~\ref{sec:t0}, we use the multi-wavelength observations and physical arguments to estimate when the associated gamma-ray burst happened. Using this estimated time, we fit the data with a structured-jet model and cocoon model. We introduce our structured-jet and cocoon models in Sec.~\ref{sec:afterglow} and perform Bayesian model selection to identify the more likely scenario. 
We explore the dirty fireball hypothesis by estimating the Lorentz factor in Sec.~\ref{sec:dirtyfireball}. 
We perform a sub-threshold search in \textit{Fermi} data and discuss the efficiency of the unobserved prompt gamma-ray emission in Sec.~\ref{sec:promptemission}.
We discuss the implications of our results and conclude in Sec.~\ref{sec:implications}. 
Our analysis suggests that AT2020blt is likely an on-axis low efficiency long gamma-ray burst. 
The lack of gamma-ray radiation can be attributed to the low radiative efficiency of AT2020blt. 
We find that the non detection in gamma rays with \konus{} and \textit{Fermi}, implies AT2020blt has a radiative efficiency $\lesssim 0.3-4.5\%$.
% , lower than up to $99.5\%$ of the observed gamma-ray burst population~\citeg{Racusin2011}
%%%%%%%%%%%%%%%%%%%%%%%%%%%%%%%%%%%%%%%%%%%%%%%%%
\section{Estimating the burst time}\label{sec:t0}
AT2020blt was first observed in the \textit{r} band on 28 January 2020, by the Zwicky Transient Facility~\citep{ztf_paper}. Follow-up observations provided detections in \textit{g} and \textit{i} bands along with detections in radio and X rays~\citep{ho20, singer20}. No coincident gamma-ray trigger was found~\citep{konus_upperlim, ho20}. The lack of a gamma-ray trigger implies that we do not know the burst time $t_{0}$, which is critical for discerning physics and testing the various hypotheses in detail. 

\citet{ho20} estimated the burst time by fitting a broken power-law simultaneously to the $r$, $g$ and $i$ band data assuming constant colour offsets. This allowed them to estimate $t_{0}$ as January 28.18. However, by fitting for $t_{0}$ with a broken power-law, \citet{ho20} have estimated the peak time of an afterglow $t_{\rm peak}$. 
For a typical on-axis relativistic jet, this peak time is likely not significantly different to $t_0$. However, for a mildly relativistic jet viewed on-axis, $t_{\rm peak}$ can be up to a few days after $t_0$~\citeg{saripiran99}, while for an off-axis system, the peak time could be several months after $t_0$~\citeg{granot02}. 

To investigate the reason for the lack of observed prompt gamma-ray emission with detailed afterglow models, we must first estimate the burst time more robustly. 
For a relativistic outflow viewed on-axis the peak timescale in optical is the deceleration timescale~\citeg{saripiran99, Chevalier2000}
\begin{equation}\label{eq:tdec}
% t_{\rm dec} \approx \unit[90]{s}~(1 + z) \left(\frac{E_{k}}{\unit[10^{50}]{erg}}\frac{\unit[10^{-2}]{cm^{-3}}}{n_{\rm ism}}\right) ^{1/3}\left(\frac{\Gamma_0}{100}\right)^{-8/3}.
t_{\rm dec} \approx \unit[10]{s}~\left(1+z\right) \left(\frac{E_{k}}{\unit[10^{52}]{erg}}\frac{1}{A_{*}}\frac{1}{\Gamma_{0}^{4}}\right),
\end{equation}
for a stellar wind medium, as expected for long gamma-ray bursts. Here, $E_{k}$ is the kinetic energy of the outflow, $z$ is the redshift, $A_{*}$ is wind parameter typically of order unity~\citep{Chevalier2000}, and $\Gamma_0$ is the initial Lorentz factor of the outflow. 
For an on-axis observer, the optical lightcurve is well modelled as a rising power-law till $t_{\rm dec}$~\citeg{saripiran99}.

The location of AT2020blt was observed prior to the first detection on January 27th with an $r$ band upper limit of $> \unit[21.36]{mag}$. We use this non-detection combined with a rising power-law on the observed flux
\begin{align}\label{eq:predeclc}
F = A \left(t - t_0\right)^m & \lesssim t_{\rm dec}, 
\end{align}
which describes the pre-deceleration physics, to estimate $t_0$ and $t_{\rm dec}$. 
We use broad uninformative priors on $t_0$, $t_{\rm dec}$ and fix $A$ to ensure the pre-deceleration power-law smoothly connects with the observations. If AT2020blt is observed on-axis, the pre-deceleration behaviour must explain the non-detection on January 27th and the afterglow flux at the start of the observations.
The prior on $m$ is informed by physics and depends on whether the forward or reverse shock dominates, or on the thickness of the shells launched in the burst. To minimise the effect of the prior, we choose a broad uniform prior from $0.5-20$, which covers all possible scenarios ranging from a forward or a reverse shock and a combination~\citeg{zhang_book} and the possibility of different stratification in the environment. 
We note that typical observed pre-deceleration behaviour follows $m \approx 3$~\citeg{zhang_book}. 
We examine whether the pre-deceleration power law would produce a signal above the upper limit on 27 January and explain the peak flux at the time of the first detection, ruling out the parameter space that violates this constraint. 

\begin{figure}
    \includegraphics[width=0.5\textwidth]{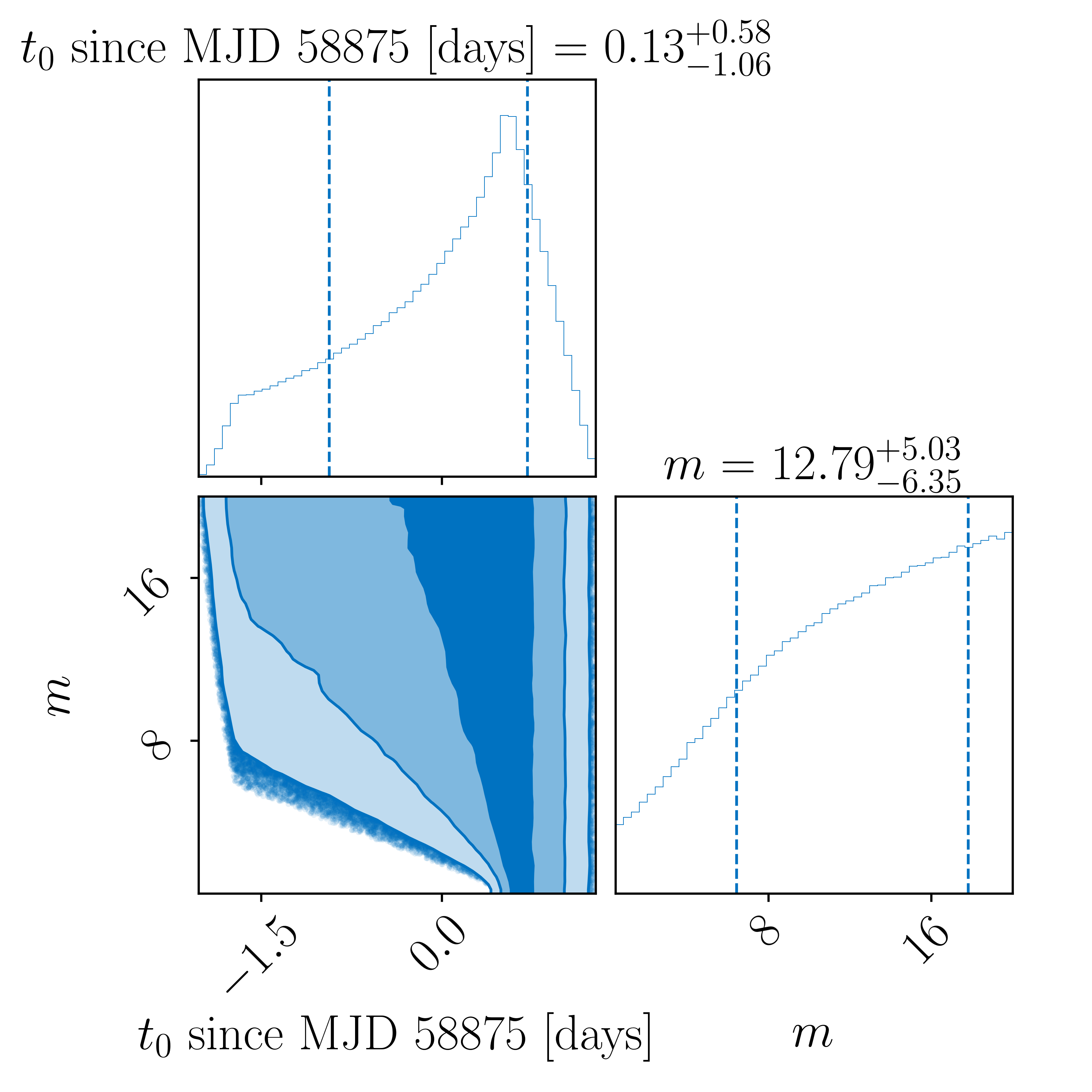} \hspace{-.4cm}
\caption{Posterior distribution for the burst time $t_{0}$ since MJD 58875 i.e., 27 January 2020 and $m$ the pre-deceleration index. The uncertainties are the $68\%$ credible interval. The shaded contours indicate the $1-3 \sigma$ credible intervals. We note that MJD 58875 is arbitrarily chosen to show the posterior distribution more clearly.}
\label{fig:t0corner}
\end{figure}

The above analysis implies that if AT2020blt is observed on-axis (i.e., the assumption of pre-deceleration is correct) then we measure $t_0 = \unit[\tstart]{MJD}$ (i.e., Jan $27.1^{+0.6}_{-1}$). 
This initial analysis can already offer some clues into the nature of AT2020blt. For shallow indices $m \lesssim 3$, as would be expected for a \cocoon{} or a relativistic jet where the forward shock dominates the pre-deceleration physics, a start time before $\sim 27$ January can not explain the non-detection. 
While for steeper indices $m \gtrsim 3$ (i.e., where reverse shock emission dominates), the range of $t_0$ is significantly broader, potentially as early as $\sim 26$ January as the rapid rise can accommodate both the non-detection and the peak observations. We estimate the deceleration (peak) time to $t_{dec} = \unit[\tp]{MJD}$ i.e., Jan $28.1^{+0.1}_{-0.2}$, $\approx 5$ hours before the first observation. Our posterior on $t_0$ and $m$ is shown in Fig~\ref{fig:t0corner}. 

The above deceleration analysis can also be extended to an off-axis observer. We emphasise that this latter physics is a consequence of relativistic beaming and not deceleration. A relativistic jet viewed off-axis will also rise with $m \gtrsim 3$. However, if the jet has an extended structure such that a significant fraction of the jet energy covers the observers' line of sight, the rise will be much shallower~\citeg{nakar19}.
This suggests that if AT2020blt was an off-axis afterglow, the observer must satisfy $\theta_{\rm obs} > 1/\Gamma$~\citeg{nakar19} at the time of the non-detection or for the gamma-ray burst to occur after the non-detection on 27 January. These two conditions combined with the steep late-time decay of AT2020blt suggests an off-axis interpretation is unfeasible. We note that the above analysis may not be valid for complex jet structures where there are two peaks in the lightcurve due to deceleration of material along the line of sight or other effects due to jet structure that produce a negative temporal slope even with relativistic beaming~\citep{Beniamini2020}. We explore the off-axis interpretation in more detail in Appendix A.
%%%%%%%%%%%%%%%%%%%%%%%%%%%%%%%%%%%%%%%%%%%%%%%%%
\section{Afterglow constraints}\label{sec:afterglow}
If AT2020blt is the afterglow of a typical gamma-ray burst, the absence of observed gamma-ray emission immediately points towards two hypotheses: either the observer was off-axis, and therefore the prompt gamma-ray emission was missed due to relativistic beaming~\citep[e.g.,][]{granot02}, or the jet did not successfully break through the ejecta and cocoon emission was responsible for producing the afterglow~\citeg{nakar17_cocoon}.  
We first test both these hypotheses by fitting the broadband afterglow to a power-law structured jet and cocoon model. We note that it is also possible that the gamma-ray emission was missed, and we explore this hypothesis later in Sec.~\ref{sec:promptemission}.

The power-law jet model is an angular structured jet with an energy distribution defined as, \begin{equation}\label{eq:jetstructure}
E(\theta_{\mathrm{observer}})=E_{\textrm{iso}}\left(1+\frac{\theta_{\mathrm{observer}}^{2}}{\beta \theta_{\mathrm{core}}^{2}}\right)^{-\beta / 2}.
\end{equation}
Here, $\beta$ is the exponent dictating the slope of the power-law jet structure, $\theta_{\mathrm{observer}}$ is the observers viewing angle, $\theta_{\mathrm{core}}$ is the half-width opening angle of the jet core. 
The Lorentz factor of the jet is proportional to $E({\theta_{\textrm{observer}}})^{1/2}$ and $E_{\textrm{iso}}$ is the on-axis isotropic equivalent energy.
The \cocoon{} model is a spherical outflow with velocity stratification with an energy distribution, 
\begin{equation}\label{eq:cocoon}
E(u)=E_{0}\left(\frac{u}{u_{\max }}\right)^{-k}.
\end{equation}
Here, $u$ is the dimensionless four-velocity, $E_{0}$ is the kinetic energy of the fastest material, and $k$ is the power-law index. 
Both the \cocoon{} and structured jet outflows interact with the surrounding interstellar medium accelerating a fraction of electrons, $\xi_{n}$, with a fraction of the total energy, $\epsilon_{e}$, and the fraction of the energy in the magnetic field, $\epsilon_b$. The radiation produced by these electrons is responsible for the observed broadband afterglow.

We fit the multi-wavelength flux density data of AT2020blt (including the upper-limits) using the power-law structured jet and \cocoon{} model described above implemented in \afterglowpy{}~\citep{afterglowpy} and a Gaussian likelihood. We set broad priors on all $15$ parameters for each models. 
We include both synchrotron and inverse Compton emission and account for potential host galaxy extinction. 
We estimate the uncertainty on each data point as the quadrature sum of the measurement uncertainty reported by~\citet{ho20} and an estimated systematic uncertainty which we model. Our prior on this systematic uncertainty is uniform from $10^{-4}-10^{-3}~\unit{mJy}$. The values being motivated by the flux data.
We infer the parameters of the system using \bilby~\citep{bilby} and the \dynesty~sampler~\citep{dynesty}.  
\begin{figure*}
  \begin{tabular}{cc}     
        \includegraphics[width=1.0\textwidth]{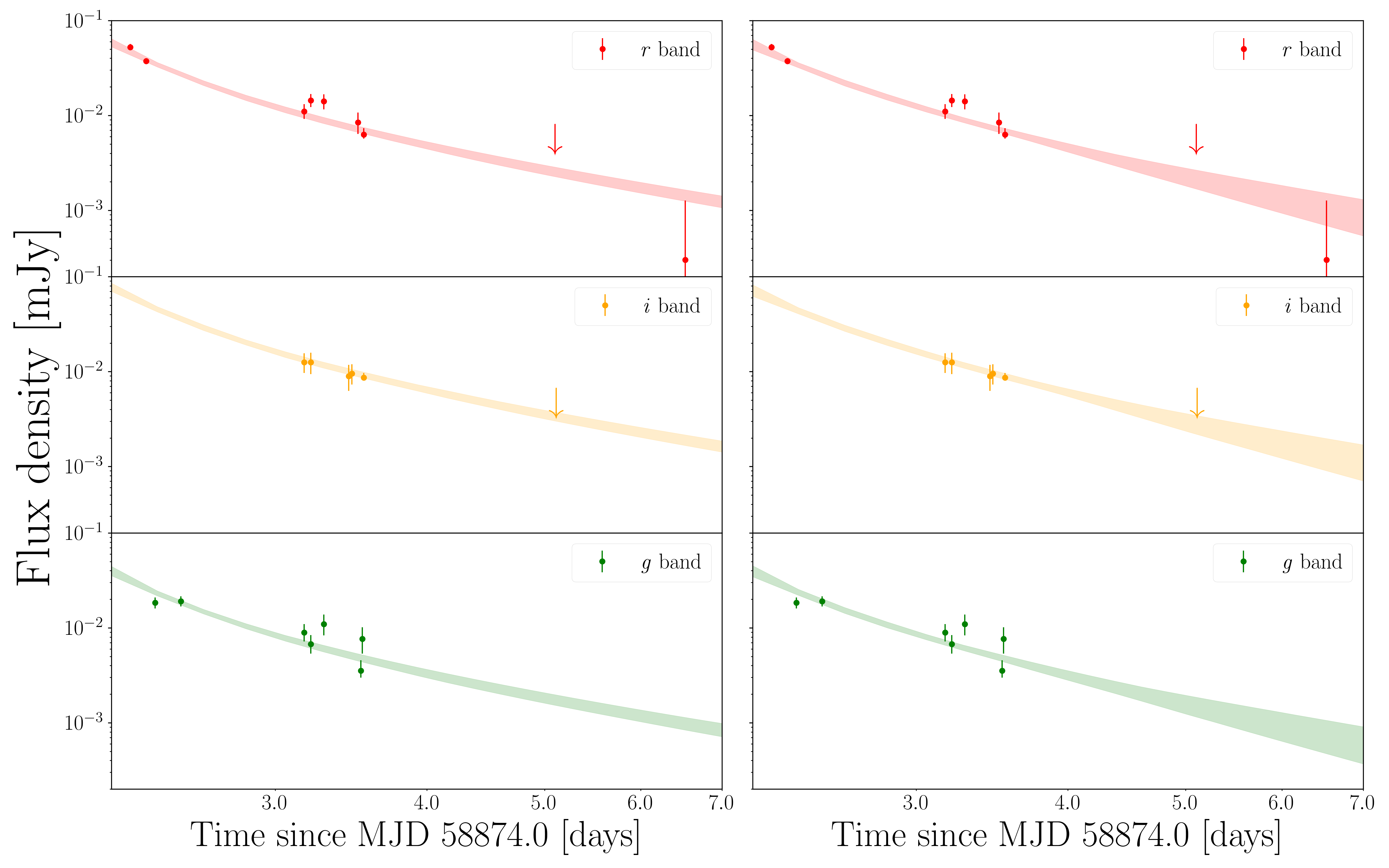} \hspace{-.4cm}
  \end{tabular}
  \caption{Data for AT2020blt in $r$, $g$, and $i$ band, with arrows indicating upper limits. In the left panel we show the $95\%$ credible interval predicted lightcurves from our posterior samples for the cocoon model. Similarly, the right panels show the $95\%$ credible interval predicted lightcurves for the power-law jet model. The errors on the data are the measurement uncertainty reported by~\citet{ho20} combined with our modelled uncertainty for each model. We note that MJD 58874 is arbitrarily chosen to show the posterior distribution more clearly. We also note that only $r$ band data is shown here but the fits are performed on the full multi-wavelength data.} 
\label{fig:splandcocoon}
\end{figure*}
Our fits to the broadband data are shown in Fig~\ref{fig:splandcocoon}. The full list of parameters and their respective priors for the structured-jet model are shown in Table.~\ref{table:priors}, while Fig.~\ref{fig:allcorners} shows the posterior distribution. We note that we do not fit the X-ray data as it is computationally expensive to simultaneously fit the flux density and integrated flux data with our model. However, after fitting the flux density data we ensure that our result is consistent with the X-ray observations. Excluding X-ray data from our fit may also be necessary as \afterglowpy{} does not incorporate synchrotron self-Compton emission which may lead to biased estimation of parameters if included when fitting the multi-wavelength data~\citeg{Nakar2009}.

We find that the \cocoon{} model does not explain the $r$-band data at late times well, while the structured-jet model can successfully explain all the observations. In particular, the last data point is a two sigma outlier from the posterior prediction for the \cocoon{} model.
Furthermore, given the pre-deceleration power-law exponent of a \cocoon{} is $m \lesssim 3$, to explain the observations with a \cocoon{} the associated gamma-ray burst must have occurred after the non-detection on 27 January. 
This requires the \cocoon{} emission to rise rapidly to explain the observations on 28 January, which is difficult~\citeg{nakar17_cocoon}.
The rapid decay of AT2020blt is also difficult to expect from a \cocoon{}. \citet{ho20} measure the temporal decay of AT2020blt at late-times as $\sim -2.6$, and such a rapid decay is difficult to expect with a \cocoon{} for a broad range of parameters~\citep{lamb18_gw170817, Troja2018}.

More quantitatively, we perform Bayesian model selection between the two hypotheses. 
Assuming both models are equally likely a priori, the structured-jet model is $\bfsplcocoon$ times more likely than the \cocoon{} interpretation, favouring the hypothesis that an ultra-relativistic structured jet broke out of the ejecta and later interacted with the surrounding environment to produce AT2020blt. 

Although we focus here on the power-law structured jet model for simplicity, we find a similar preference for other jet models, including a Gaussian structured jet and a top-hat jet with Bayes factors of $\bfgaussiancocoon$ and $\bftophatcocoon$ in favour of the jet hypothesis respectively.
We emphasise that these Bayes factors are predicated on the assumption that AT2020blt was a gamma-ray burst i.e., we are comparing the cocoon and structured jet hypotheses given AT2020blt is a gamma-ray burst. In Sec.~\ref{sec:implications}, we speculate that it is possible, although unlikely, that AT2020blt may have a different origin, in which case this analysis would not be valid. 

As we are also modelling the noise, another way to compare the model fit is by the size of the estimated noise. We find $\sigma = \splsigma$ and $\sigma = \cocoonsigma$ for the power-law structured jet and \cocoon{} models respectively, i.e., to fit the data with the \cocoon{} model, we need the data to be noisier, suggesting that the structured-jet model is favoured over the \cocoon{}. 

The inability to explain the observations with a \cocoon{} while successfully explaining the observations with a structured jet suggests AT2020blt had a successful jet breakout, as typical for most observed gamma-ray bursts. Therefore, the lack of observed prompt gamma-ray emission could be a consequence of relativistic beaming, i.e., that we observed AT2020blt outside the ultra-relativistic core. 
We explore this hypothesis in detail in Appendix~\ref{app:offaxis} finding that we can not fit the data well, with a Bayes factor of $\bfsploffaxis$ in favour of the on-axis hypothesis. This is likely due to the sharp rise required to explain the non-detection and first observation around 28 January while also explaining the subsequent steep decay thereafter, which is difficult to expect for an observer located off-axis~\citeg{granot02, nakar19}. 

\begin{figure}
    \includegraphics[width=0.5\textwidth]{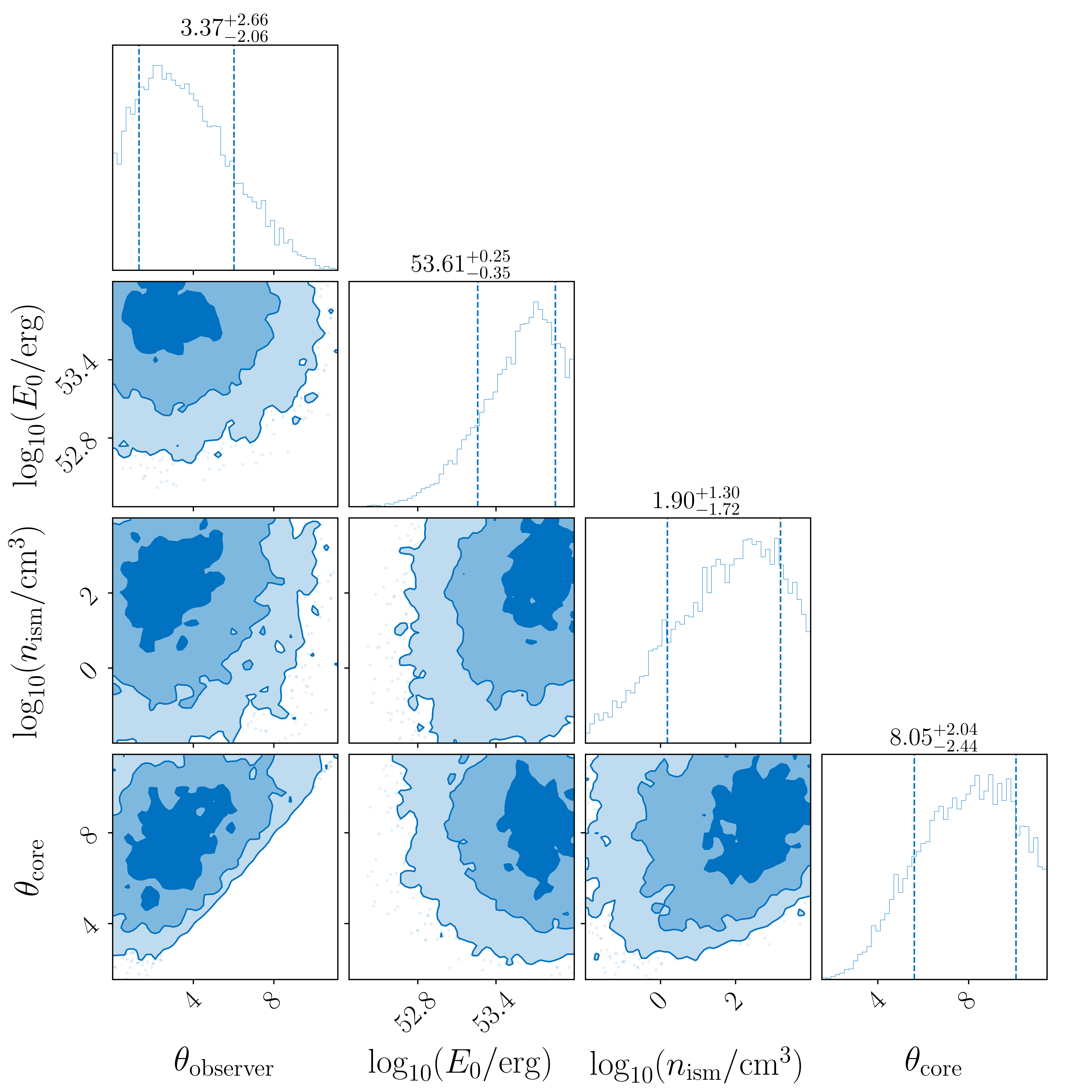} \hspace{-.4cm}
\caption{Posterior distributions for the structured jet model observers viewing angle $\theta_{\textrm{observer}}$, the half-width jet core $\theta_{\textrm{core}}$, the on-axis isotropic equivalent energy $E_{\textrm{iso}}$, and the ambient interstellar medium density $n_{\textrm{ism}}$ for AT2020blt. The shaded contours indicate the $1-3 \sigma$ credible intervals.}
\label{fig:corner}
\end{figure}

In Fig.~\ref{fig:corner}, we show the one and two-dimensional posterior distributions for the observers viewing angle $\theta_{\textrm{observer}}$, the half-width jet core $\theta_{\textrm{core}}$, the isotropic equivalent energy $E_{\textrm{iso}}$, and the ambient interstellar medium density $n_{\textrm{ism}}$. 
Our measurement of the ultra-relativistic core of the jet is $\theta_{\textrm{core}} = \thetacore$,  while $\theta_{\textrm{observer}} = \thetaobs$, implying that we observed AT2020blt on-axis much like most other gamma-ray bursts. The credible intervals are $68\%$ unless specified otherwise. We also measure the on-axis isotropic equivalent energy $\log_{10} (E_{\textrm{iso}}/{\unit{erg}}) = \eiso$, and the ambient interstellar medium density $\log_{10} (n_{\textrm{ism}}/{\textrm{cm}^{-3}}) = \nism$ consistent with the population of other gamma-ray bursts~\citep[e.g.,][]{Nava2014, beniamini15, wang15, fong15}. Our constraint on the fraction of energy in electrons $\log_{10} \epsilon_{e} = -1.1 \pm 0.3$ is also consistent with measurements from radio afterglows of gamma-ray bursts~\citep{Nava2014, Beniamini2017}.

These measurements suggest that AT2020blt is consistent with being the afterglow of a relatively typical long gamma-ray burst where the observer is located within the ultra-relativistic core. This implies that the non-detection of prompt gamma-ray emission cannot be due to relativistic beaming i.e., we did not observe AT2020blt off-axis. 

Our results confirm that assuming AT2020blt was related to the gamma-ray burst phenomena, the data of AT2020blt is best explained as the afterglow produced by an on-axis jet that successfully broke out of the ejecta. This implies that we can rule out the non-detection of prompt gamma-ray emission due to relativistic beaming or that the jet that produced AT2020blt did not successfully break out of the ejecta. We now explore whether the jet launched in AT2020blt could produce prompt gamma-ray emission.  
%%%%%%%%%%%%%%%%%%%%%%%%%%%%
\section{Lorentz factor}\label{sec:dirtyfireball}
As discussed above, gamma-ray production requires a relativistic jet to alleviate the compactness problem~\citep{ruderman75}. Naturally, this implies that if a jet breaks out of the ejecta and it is not sufficiently relativistic (for potential explanations of non-relativistic jets see e.g.,~\cite{huang02}), the jet will not produce detectable prompt gamma-ray emission. 

In Sec.~\ref{sec:afterglow}, we showed that AT2020blt is on-axis and likely successfully launched a jet that broke out of the ejecta. Here, we explore if the jet was above the prompt emission threshold through back-of-the-envelope estimates and detailed fitting.

The threshold for producing prompt gamma-ray emission is typically assumed to be $\Gamma_0 \gtrsim 100$, with jets with Lorentz factors below this threshold referred to as `dirty' fireballs or failed gamma-ray bursts~\citep{huang02, rhoads03}. 
However, indirect measurements of the Lorentz factor as low as $\Gamma_0 \sim 20$ have been made for jets following some successful gamma-ray bursts~\citep[e.g.,][]{ghirlanda18} with weak prompt emission. 
Accurately defining the gamma-ray prompt emission threshold requires knowing the radius at which prompt emission is produced, through what mechanism, and the fraction of photons above the pair production threshold in the co-moving frame. We do not know any of these constraints. However, all of these unknowns serve to lower the threshold $\Gamma_0$ value for producing prompt gamma-ray emission. As such, we take the conservative value of $\Gamma_0 \sim 100$ as the gamma-ray emission threshold, with $\Gamma_0 \sim 20$ serving as the absolute lower limit. We note that the latter threshold comes from assuming a wind-like interstellar medium.

We estimate the Lorentz factor by fitting the same power-law structured jet model described earlier with a finite deceleration radius allowing us to estimate the Lorentz factor. 
Our posterior on the Lorentz factor is $\Gamma_0 = \lorentz$. 
The probability of having the Lorentz factor lower than the prompt emission threshold ($\Gamma_0 \lesssim 100$) is $\lesssim \percentagehundred\%$ and $\lesssim \percentagetwenty\%$ for $\Gamma_0 \lesssim 20$. 
However, we note that because AT2020blt is missing early time data, our result is strongly dependent on the prior we assume for $\Gamma_0$ and implicitly on the full distribution of $t_0$ derived in Sec.~\ref{sec:t0}. 
For the analysis above, we place a uniform prior on $\Gamma_0$ between $1-1000$. To mitigate this dependency on the prior, we also estimate the Lorentz factor through back-of-the-envelope arguments.

The Lorentz factor can be estimated by measuring the afterglow onset time, also referred to as the deceleration timescale $t_{\rm{dec}}$. For an on-axis observer, the deceleration timescale is the peak of the optical lightcurve, with the relativistic jet starting to decelerate on this timescale. This allows us to place a lower limit on $\Gamma_0$. 
The Lorentz factor is related to the deceleration timescale and weakly to the interstellar medium density and jet energy. We can approximate this relationship by~\citep[e.g.,][]{saripiran99, nakar07}
\begin{equation}\label{eq:afterglowonset}
\Gamma_0 \approx 40\left(\frac{E_{k, \rm{iso}, 50}}{n_{\rm{ism}}}\right)^{1/8} \left(\frac{100 (1 + z)}{t_{\rm{peak}} - t_{0}}\right)^{3/8}.
\end{equation}
Here, $E_{k, \rm{iso}, 50}$ is the isotropic equivalent kinetic energy in units of $10^{50}$ erg, which we measure from the fitting of the afterglow, and $z$ is the redshift. 
We do not observe this peak in AT2020blt but our analysis from Sec.~\ref{sec:t0} provides a conservative estimate of $t_{\rm peak} = \unit[\tp]{MJD}$. 
With this estimate for $t_{\rm {peak}}$ we can use Eq.~\ref{eq:afterglowonset} and the derived values of $E_{k, \rm{iso}}$ and $n_{\rm{ism}}$ from Sec.~\ref{sec:afterglow} to set a lower limit on the Lorentz factor. We emphasise that the derived estimates of $E_{\rm{iso}}$ and $n_{\rm{ism}}$ are robust to the choice of the prior. Moreover, given the weak dependence of $\Gamma_0$ to $E_{k, \rm{iso}}$ and $n_{\rm{ism}}$, the predominant source of uncertainty is from the estimate of $t_0$ and $t_{\rm peak}$ itself.
Taking the median values of our estimated parameters suggests, conservatively, $\Gamma_0 \gtrsim 15$, indicating that AT2020blt may have successfully produced prompt gamma-ray emission. 
The full distribution of $t_0$, $t_{\rm{peak}}$ etc implies a range on $\Gamma_0$ from $4-400$.

We note that if AT2020blt was a `dirty' fireball or failed gamma-ray burst, the threshold for producing gamma-rays would be higher. This is due to gamma-ray photons being produced when the jet is still optically thick and subsequently reabsorbed into the outflow, raising the kinetic energy of the outflow~\citeg{lamb_kobayashi16}. A scaling for this threshold is $\Gamma_{0} \sim 16 (E_{k, 50})^{0.15}$, which is weakly dependent on the efficiency in turning the gamma-ray energy into kinetic. For our estimated kinetic energy from Sec.~\ref{sec:afterglow}, this implies $\Gamma_0 \sim 56$ for producing gamma-ray emission, above our conservative estimate. However, our estimated Lorentz factor above is conservative and if AT2020blt is not a dirty fireball then this analysis does not hold. We therefore work with the observationally supported threshold where a jet with $\Gamma_0 \sim 20$ produced a successful gamma-ray burst.

%If instead, we assume a uniform distribution on $t_{\rm{dec}}$ with an upper limit of $0.1$ days and take into account the uncertainty on $E_{k, 50}$ and $n_{\rm{ism}}$ from our afterglow analysis, we estimate $\Gamma_0 ~ 194^{+621}_{-147}$, suggesting conservatively, that there is only a $34\%$ probability 
Our analysis indicates that AT2020blt is likely not the afterglow from a `dirty' fireball and that the jet that broke out of the ejecta was likely above the threshold for producing prompt gamma-ray emission. 

The lack of observed gamma-rays is therefore puzzling. As we show in Sec.~\ref{sec:afterglow}, we observed AT2020blt on-axis, and therefore, relativistic beaming cannot explain the non-detection of prompt gamma-rays. Furthermore, our analysis indicates that AT2020blt can not be from a \cocoon{}. Having ruled both these hypotheses out, we now turn to look at the prompt emission efficiency itself.

\section{Prompt emission efficiency}\label{sec:promptemission}
Despite over three decades of observations, we still do not understand how prompt gamma-ray emission is produced. Given this uncertainty, we do not have a robust generative model to predict the energetics of the prompt gamma-ray emission. 
Given the lack of a model, a common approach in the field is to compare the energetics of the prompt and afterglow phases and compute a radiative efficiency $n_{\gamma}$
\begin{equation}\label{eq:efficiency}
n_{\gamma} = \frac{E_{\gamma, \rm{iso}}}{E_{k, \rm{iso}} + E_{\gamma, \rm{iso}}},   
\end{equation}
where $E_{\gamma, \rm{iso}}$ is the observed isotropic energy in gamma-rays. This efficiency has been calculated for a large catalogue of long and short gamma-ray bursts~\citep{wang15, fong15} using various techniques that have their own associated problems, most notably, fixing the energy in the magnetic-field $\epsilon_{b}$, rather than marginalising over the uncertainty in this parameter. The efficiencies have a broad distribution ranging from $\sim 1$ to $\sim 90$\%. 
In principle, the radiative efficiency can offer a clue into the prompt emission mechanism. However, this is fraught with uncertainties due to detector selection effects, uncertain physics~\citep[e.g.,][]{lloyd04, zhang07} and likely incorrect modelling assumptions such as ignoring synchrotron energy losses due to synchrotron self-Compton emission, which is inconsistent with recent constraints~\citep{Beniamini2016}.

\cite{ho20} searched for sub-threshold triggers in \textit{Fermi}~\citep{fermi09} and \konus~\citep{konus-wind} finding no potential counterpart. Based on IPN observations, \cite{ho20} placed an upper limit on the gamma-ray energy of $E_{\gamma} \lesssim 7 \times 10^{52}~\unit{erg}$ by taking a nominal fluence threshold of $10^{-6}$ erg/cm$^{2}$. 
Given our estimate for the kinetic energy through the afterglow fitting (see Sec.~\ref{sec:afterglow}) this implies $n_{\gamma} \lesssim \efficiencyipn$. However, this threshold is potentially conservative. A search for coincident gamma-rays from \konus{} has instead set a deeper upper limit on the fluence of $6.1\times 10^{-7}$ erg/cm$^{2}$~\citep{konus_upperlim}. With our estimate of the kinetic energy, this implies $n_{\gamma} \lesssim \efficiencykonus$ for gamma-rays not to be observed by \konus{} or a gamma-ray energy $E_{\gamma} \lesssim \unit[1.2\times 10^{52}]{erg}$. 
We note that the latter upper limit is set for a typical gamma-ray burst spectrum lasting \unit[2.9]{s} and may not apply to AT2020blt.

To augment this upper limit, we perform a sub-threshold search of \textit{Fermi} data from $\unit[1]{min}$ before the first non-detection (i.e., on January 27.54) to $\unit[2]{min}$ after the first detection (i.e., on January 28.28). Unlike \konus{} which did not have any interruptions to observations, \textit{Fermi} observations were periodically interrupted due to occultation by Earth and \textit{Fermi} passing through the South Atlantic Anomaly. In our given search window, \textit{Fermi} was observing the location of AT2020blt $\sim 64\%$ of the total time making it plausible that a gamma-ray burst weaker than \konus{} was missed. However, we note that around our most probable start time $t_0=\unit[58875.53]{MJD}$, \textit{Fermi} did not have any interruptions.
The constraints on the gamma-ray prompt emission from \textit{Fermi} are more sensitive than \konus{}, with an upper limit on $E_{\gamma, \rm{iso}}$ from $1-6 \times 10^{51}~\unit{erg}$ depending on the spectrum of the source and spectral template used in the search. This corresponds to an upper limit on $n_{\gamma} \lesssim 1.3\%$ and $\lesssim 0.3\%$ respectively.
For details of the sub-threshold search, source spectrum and templates, we refer the reader to~\citet{Blackburn2015, Goldstein2016, goldstein19}.
Taking the lower value implies that for AT2020blt to not produce detectable prompt emission, the radiative efficiency must have been lower than $\efficiencyfermi$, challenging several prompt emission models such as photospheric emission~\citep[e.g.,][]{lazzati13} and magnetic field dissipation~\citep[e.g.,][]{zhang11}, but plausible for internal shock models~\citep[e.g.,][]{Kobayashi1997, Daigne1998, kumar99}. Such a low efficiency is only measured in one other gamma-ray burst, GRB190829A which had an efficiency of $\sim 0.1\%$~\citeg{Salafia2021}.
 
\begin{figure}
    \includegraphics[width=0.5\textwidth]{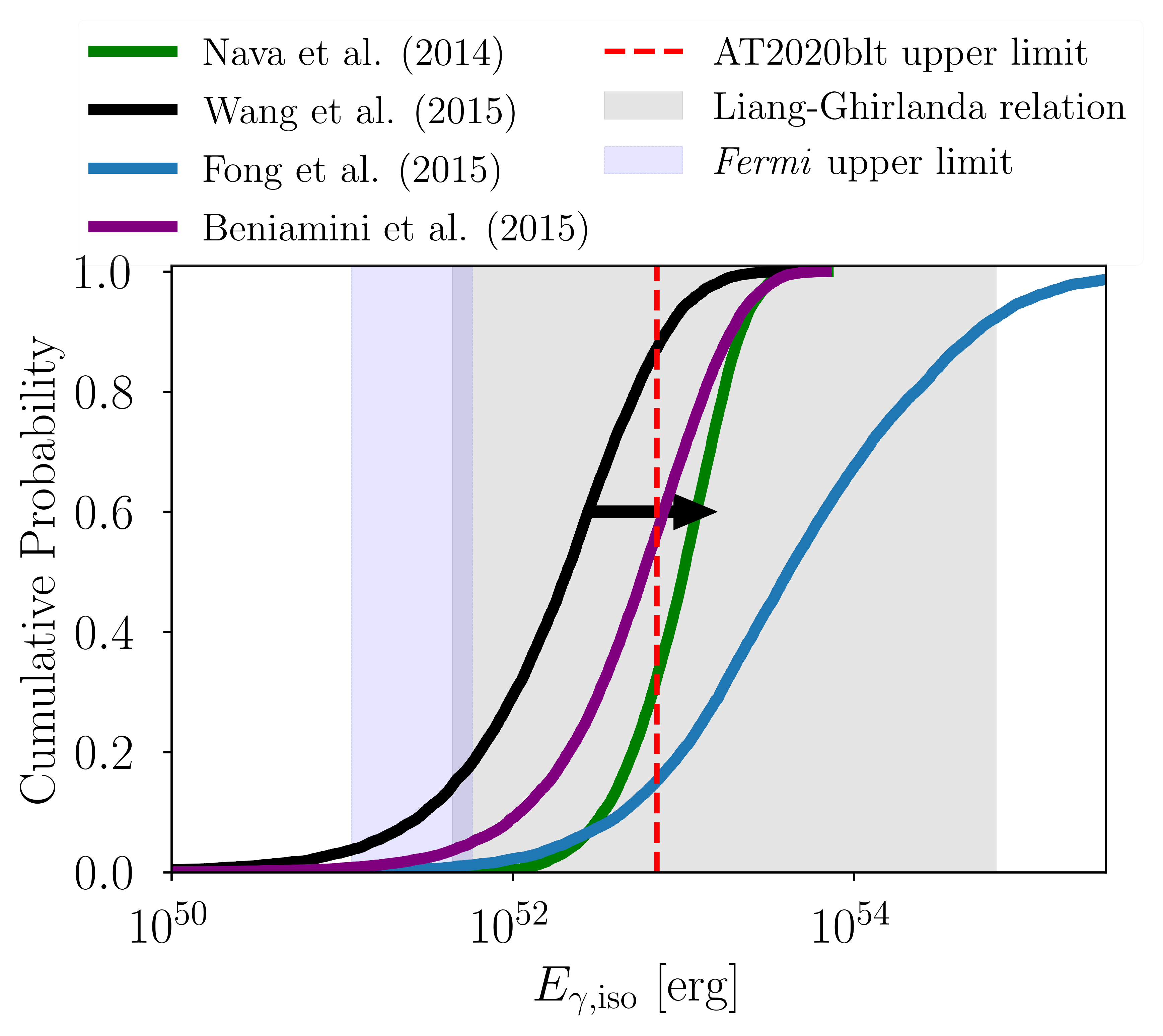} \hspace{-.4cm}
\caption{Cumulative probability distributions of the isotropic equivalent energy in gamma-rays $E_{\rm{\gamma, iso}}$ of AT2020blt. The green, black, blue, and purple curves are the cumulative distributions of gamma-ray burst energies for AT2020blt calculated using the distribution of efficiencies derived from the \citet{Nava2014}, \citet{wang15}, \citet{fong15}, and \citet{beniamini15} analyses, respectively. The black arrow indicates that the distribution derived in~\citep{wang15} is likely a lower limit. The red line indicates the conservative upper limit for AT2020blt based on IPN. observations set by~\citet{ho20}, while the blue shaded region shows the upper limit from a sub-threshold search we perform with \textit{Fermi}. The grey band is the prediction for the gamma-ray energy from Liang-Ghirlanda relation.} 
\label{fig:eiso_cdf}
\end{figure}

In Fig.~\ref{fig:eiso_cdf}, we show these upper limits on the isotropic energy in gamma-rays for AT2020blt. The green, black, blue, and purple curves are the cumulative distributions of gamma-ray burst energies for AT2020blt calculated using the distribution of efficiencies derived from the \cite{Nava2014}, \cite{wang15}, \cite{fong15}, and \cite{beniamini15} analyses, respectively. The latter three analyses all calculate efficiencies for a sample of gamma-ray bursts with various assumptions. In particular, the distribution of efficiencies from \cite{beniamini15} is calculated using the GeV flux rather than the X-ray flux as the former is likely a better proxy for the true kinetic energy of the afterglow. The green curve, however, is calculated assuming the average efficiency of gamma-ray bursts is $0.2$, consistent with other estimates~\citeg{Beniamini2016} and using the constraint that the distribution of efficiencies is narrow as determined by~\citet{Nava2014}. The true average efficiency of gamma-ray bursts is unknown which would shift the green curve left or right for lower and higher average efficiencies respectively. 
We note that as \cite{wang15} fix $\epsilon_b = 10^{-5}$ for all gamma-ray bursts, the efficiency derived in their sample are systematically biased to low values and serve as a lower limit on the observed population. 
A similar bias due to fixing of parameters is present in other analyses~\citep[see for e.g.,][]{Racusin2011, fong15} and therefore comparisons to populations should be done cautiously.

The non-detection by \konus{} and \textit{Fermi} implies that if AT2020blt produced prompt emission, the isotropic energy released in gamma-rays must have been lower than $\sim 1\times10^{51} \unit{erg}$, an efficiency weaker than $\sim 99.9\%$ of the observed distribution of long gamma-ray bursts~\citep{fong15} or $97.5\%$ if considering the distribution derived in~\citet{wang15}. In the gray shaded region, we show the predicted isotropic gamma-ray energy through the Liang-Ghirlanda relation, $E_{\gamma} \sim 10^{52}(\Gamma_0/182)^{4}$~\citep[e.g.,][]{liang10, ghirlanda18}. Here, we use our posterior, $\Gamma_0 = \lorentz$ obtained by fitting the afterglow (see Sec.~\ref{sec:afterglow}). Given the non-detection by \textit{Fermi} and \konus{}, AT2020blt likely did not follow the Liang-Ghirlanda relation, which may suggest a different origin for this transient. However, it is worth noting that there are notable exceptions to the Liang-Ghirlanda relation, such as GRB090510~\citep{ghirlanda10}.

Our results suggest that if AT2020blt is a typical gamma-ray burst, given the afterglow modelling indicates that this event was on-axis, the prompt emission must have been low-luminosity to not be detected by \textit{Fermi} and \konus{} or missed by \textit{Fermi} and with a radiative efficiency of $n_{\gamma} \lesssim 2.8-4.5\%$. Or equivalently, weaker than up to $98.2\%$ of the population~\citep{fong15}. This low efficiency is not necessarily a problem for prompt emission models~\citep[e.g.,][]{Kobayashi1997, Daigne1998, kumar99} and might even support certain varieties of internal shock models. However, it is inconsistent with a large fraction of observed gamma-ray bursts and previously derived efficiency distributions~\citep[e.g.,][]{Racusin2011, wang15, fong15}. 
It is also plausible that AT2020blt is a failed gamma-ray burst, but to confirm this hypothesis, we need to accurately determine the threshold for generating gamma-ray emission. If the threshold is $\Gamma_{0}\sim 20$ (motivated by observations of successful gamma-ray bursts) then our most conservative analysis suggests AT2020blt is not a failed gamma-ray burst. However, if the threshold is higher, then there is some part of the parameter space where the jet launched in AT2020blt could fail to produce gamma-ray emission.
% \begin{figure}
%     \includegraphics[width=0.5\textwidth]{gbm_upper_limits_10_1000_kev.pdf} \hspace{-.4cm}
% \caption{Coverage of Fermi GBM} 
% \label{fig:fermigbm}
% \end{figure}
%%%%%%%%%%%%%%%%%%%%%%%%%%%%%%%%%%%%%%%%%%%%%%%%%
\section{Implications and conclusion}\label{sec:implications}
Gamma-ray bursts have been observed extensively for over three decades, with their lower energy broadband afterglows almost always observed in the follow-up to a gamma-ray trigger. The Zwicky Transient Facility is quickly changing this dynamic, with three afterglow-like transients already detected without a gamma-ray trigger~\citep{ho20, at2021any}, with a further four independent detections of an optical counterpart to a gamma-ray burst. Here, we have investigated the observations of AT2020blt in detail to ultimately determine why no prompt gamma-ray emission was observed?

In Sec.~\ref{sec:afterglow}, we showed that AT2020blt is best interpreted as the afterglow produced by a successful on-axis structured jet. We ruled out the hypotheses that AT2020blt was the afterglow of a \cocoon{} or an off-axis gamma-ray burst, with a \cocoon{} and off-axis jet model unsuccessful in explaining the observations.
In Sec.~\ref{sec:dirtyfireball}, we showed that the jet launched in AT2020blt was likely above the prompt emission generation threshold, i.e., the jet was relativistic enough to alleviate the compactness problem, potentially ruling out the hypothesis that AT2020blt was the afterglow of a dirty fireball. Moreover, given the rate of afterglow-like transients and gamma-ray bursts are roughly consistent, such phenomena can potentially already be ruled out~\citeg{cenko15, ho20}.
In Sec.~\ref{sec:promptemission}, we found that the non-detections of prompt gamma-ray emission in \textit{Fermi} and \konus{} imply that the prompt gamma-ray emission in AT2020blt is weaker than up to $99.5\%$ of the observed gamma-ray burst population.  

% The prompt emission we observed for GRB 170817A was likely produced through a different mechanism to other gamma-ray bursts, with cocoon-shock breakouts as one of the leading candidates~\citep[e.g.,][]{gottleib18}. However, given the large redshift of AT2020blt, a similar signature would have likely been undetectable~\citep[e.g.,][]{nakar12}. 
We also predicted the prompt emission energy through the Liang-Ghirlanda relation. If AT2020blt was consistent with this relation, the prompt emission should have been detected by both \textit{Fermi} and \konus{}, implying the gamma-ray energy generated in AT2020blt must be lower than the predictions by the Liang-Ghirlanda relation. However, we note that there are other gamma-ray burst exceptions to this relation, particularly GRB 090510~\citep{ghirlanda10}, which like AT2020blt, had weaker gamma-ray emission than expected.

Given our inadequate knowledge of prompt emission generation, the only meaningful study about the prompt emission we can do is investigate the prompt emission efficiency. Depending on the upper limits used, the radiative efficiency of AT2020blt is $n_{\gamma} \lesssim 0.3-4.5\%$, which is lower than between $86.3-99.5\%$ of the observed long gamma-ray burst population~\citep{Racusin2011}. This low efficiency strongly favours internal shocks as the likely prompt emission generation mechanism~\citeg{kumar99}. 

In addition to AT2020blt, the Zwicky Transient Facility also detected AT2021any, another potential afterglow-like transient at $z = 2.514$~\citep{at2021any, at2021any_2} which was observed to be rapidly fading after $\sim 22 \unit{mins}$ and detected without an identified gamma-ray counterpart despite coverage from multiple gamma-ray telescopes~\citep{at2021any}. 
Given that AT2021any was rapidly fading after $\sim 22 \unit{mins}$, it was also likely observed on-axis and the non-detection of gamma-ray emission could imply it also has a low prompt emission efficiency that is inconsistent with the population or probing the margins of the efficiency distribution of gamma-ray bursts~\citep{Nava2014}. 
We note that the data for AT2021any is not yet available and therefore we can not perform the detailed analyses for AT2021any we presented here.
 
The radiative efficiencies of AT2020blt and AT2021any are potentially inconsistent with (or at least in the tails of) the observed population of gamma-ray bursts. 
The capabilities of the Zwicky Transient Facility, in particular, the high cadence and large survey volume~\cite[e.g.,][]{ztf_paper} provide an opportunity to detect afterglow transients without a gamma-ray counterpart. 
This implies that the detection of transients such as AT2020blt and AT2021any are independent of gamma-ray observatories' observational biases. 
Therefore, it is conceivable that AT2020blt and AT2021any are the afterglows produced by the low-luminosity gamma-ray bursts that have so far been missed entirely (or less frequently observed) due to a Malmquist bias associated with gamma-ray observatories.
Future all-sky gamma-ray detectors can avoid this bias and test whether these transients are genuinely part of the cosmological gamma-ray burst population or a distinct new class. 
Furthermore, new optical telescopes like the Vera Rubin observatory~\citeg{lsst} will be capable of finding more transients like AT2020blt to deeper magnitudes and provide stringent constraints on various parameters to test the nature of these transients.

We also note that the potential inconsistency of the radiative efficiencies of AT2020blt and AT2021any with the observed population may be a product of inadequate modelling and systematic biases in previous studies into the radiative efficiencies of gamma-ray bursts. 
Inferring prompt emission efficiencies requires robustly determining the kinetic energy in the afterglow, which requires a good understanding of the jet structure (at least for off-axis sources). 
Furthermore, in previous analyses, the estimations have often been done by fixing the energy fraction in the magnetic field, $\epsilon_{b}$ and the participation fraction, $\xi_{n}$, which significantly underestimates uncertainties at best or leads to biases at worst. This motivates the need for detailed afterglow modelling on individual gamma-ray bursts and entire populations with a detailed treatment of observational selection effects~\citep[e.g.,][]{mandel19}. 

It is also worth considering\textemdash{}albeit cautiously\textemdash{}whether AT2020blt and AT2021any are afterglows of typical gamma-ray burst progenitors at all. Instead, they may be the product of a different phenomenon or a subclass of typical gamma-ray burst progenitors that can launch a mildly relativistic, ultra-low efficiency jet that interacts with the interstellar medium to produce an afterglow-like transient but does not necessarily produce prompt gamma-ray emission. If the Zwicky Transient Facility and later the Vera Rubin observatory continues to find afterglow-like transients without prompt emission, it would be intriguing to consider the host galaxy properties and the population properties of such transients. Although, as we discussed previously, it is more likely that AT2020blt and AT2021any are simply afterglows of gamma-ray bursts we previously missed due to observational biases. By virtue of opening a new window into these phenomena, it is likely that the Zwicky Transient Facility and in the future, Vera Rubin observatory will continue to find afterglow-like transients without high energy counterparts. 
%%%%%%%%%%%%%%%%%%%%%%%%%%%%%%%%%%%%%%%%%%%%%%%%%%%%%%%%%%%%%%%%%%%%%%%%%%%%%%
\section{Acknowledgments}
We are grateful to Kendall Ackley, Moritz H\"ubner, Paul Easter, and the anonymous referee for their thoughtful comments on the manuscript. N.S. is supported by an Australian Government Research Training Program (RTP) Scholarship. P.D.L. is supported through Australian Research Council Future Fellowship FT160100112 and CE170100004. P.D.L. and G.A. are supported by ARC. Discovery Project DP180103155. 
% This document contains \total{citnum} references.
%%%%%%%%%%%%%%%%%%%%%%%%%%%%%%%%%%%%%%%%%%%%%%%%%%
\section{Data Availability}
All the data used in this paper is available in \cite{ho20}.



\bibliographystyle{mnras}
\bibliography{ref}

\begin{thebibliography}{}
\makeatletter
\relax
\def\mn@urlcharsother{\let\do\@makeother \do\$\do\&\do\#\do\^\do\_\do\%\do\~}
\def\mn@doi{\begingroup\mn@urlcharsother \@ifnextchar [ {\mn@doi@}
  {\mn@doi@[]}}
\def\mn@doi@[#1]#2{\def\@tempa{#1}\ifx\@tempa\@empty \href
  {http://dx.doi.org/#2} {doi:#2}\else \href {http://dx.doi.org/#2} {#1}\fi
  \endgroup}
\def\mn@eprint#1#2{\mn@eprint@#1:#2::\@nil}
\def\mn@eprint@arXiv#1{\href {http://arxiv.org/abs/#1} {{\tt arXiv:#1}}}
\def\mn@eprint@dblp#1{\href {http://dblp.uni-trier.de/rec/bibtex/#1.xml}
  {dblp:#1}}
\def\mn@eprint@#1:#2:#3:#4\@nil{\def\@tempa {#1}\def\@tempb {#2}\def\@tempc
  {#3}\ifx \@tempc \@empty \let \@tempc \@tempb \let \@tempb \@tempa \fi \ifx
  \@tempb \@empty \def\@tempb {arXiv}\fi \@ifundefined
  {mn@eprint@\@tempb}{\@tempb:\@tempc}{\expandafter \expandafter \csname
  mn@eprint@\@tempb\endcsname \expandafter{\@tempc}}}

\bibitem[\protect\citeauthoryear{{Abbott}, {Abbott}, {Abbott}, {Acernese},
  {Ackley}  et~al.}{{Abbott} et~al.}{2017}]{abbott17_gw170817_gwgrb}
{Abbott} B.~P.,  {Abbott} R.,  {Abbott} T.~D.,  {Acernese} F.,  {Ackley} K.,
  et~al., 2017, \mn@doi [Astrophys. J.] {10.3847/2041-8213/aa920c}, \href
  {http://adsabs.harvard.edu/abs/2017ApJ...848L..13A} {848, L13}

\bibitem[\protect\citeauthoryear{{Aloy}, {Janka}  \& {M{\"u}ller}}{{Aloy}
  et~al.}{2005}]{aloy05}
{Aloy} M.~A.,  {Janka} H.~T.,   {M{\"u}ller} E.,  2005, \mn@doi [\aap]
  {10.1051/0004-6361:20041865}, \href
  {https://ui.adsabs.harvard.edu/abs/2005A&A...436..273A} {436, 273}

\bibitem[\protect\citeauthoryear{{Aptekar} et~al.,}{{Aptekar}
  et~al.}{1995}]{konus-wind}
{Aptekar} R.~L.,  et~al., 1995, \mn@doi [\ssr] {10.1007/BF00751332}, \href
  {https://ui.adsabs.harvard.edu/abs/1995SSRv...71..265A} {71, 265}

\bibitem[\protect\citeauthoryear{{Ashton} et~al.,}{{Ashton}
  et~al.}{2019}]{bilby}
{Ashton} G.,  et~al., 2019, \apjs, 241, 27

\bibitem[\protect\citeauthoryear{{Bellm} et~al.}{{Bellm}
  et~al.}{2019}]{ztf_paper}
{Bellm} E.~C.,  et~al., 2019, \mn@doi [\pasp] {10.1088/1538-3873/aaecbe}, \href
  {https://ui.adsabs.harvard.edu/abs/2019PASP..131a8002B} {131, 018002}

\bibitem[\protect\citeauthoryear{{Beniamini} \& {van der Horst}}{{Beniamini} \&
  {van der Horst}}{2017}]{Beniamini2017}
{Beniamini} P.,  {van der Horst} A.~J.,  2017, \mn@doi [\mnras]
  {10.1093/mnras/stx2203}, \href
  {https://ui.adsabs.harvard.edu/abs/2017MNRAS.472.3161B} {472, 3161}

\bibitem[\protect\citeauthoryear{{Beniamini}, {Nava}, {Duran}  \&
  {Piran}}{{Beniamini} et~al.}{2015}]{beniamini15}
{Beniamini} P.,  {Nava} L.,  {Duran} R.~B.,   {Piran} T.,  2015, \mn@doi
  [\mnras] {10.1093/mnras/stv2033}, \href
  {https://ui.adsabs.harvard.edu/abs/2015MNRAS.454.1073B} {454, 1073}

\bibitem[\protect\citeauthoryear{{Beniamini}, {Nava}  \& {Piran}}{{Beniamini}
  et~al.}{2016}]{Beniamini2016}
{Beniamini} P.,  {Nava} L.,   {Piran} T.,  2016, \mn@doi [\mnras]
  {10.1093/mnras/stw1331}, \href
  {https://ui.adsabs.harvard.edu/abs/2016MNRAS.461...51B} {461, 51}

\bibitem[\protect\citeauthoryear{{Beniamini}, {Granot}  \& {Gill}}{{Beniamini}
  et~al.}{2020}]{Beniamini2020}
{Beniamini} P.,  {Granot} J.,   {Gill} R.,  2020, \mn@doi [\mnras]
  {10.1093/mnras/staa538}, \href
  {https://ui.adsabs.harvard.edu/abs/2020MNRAS.493.3521B} {493, 3521}

\bibitem[\protect\citeauthoryear{{Blackburn}, {Briggs}, {Camp}, {Christensen},
  {Connaughton}, {Jenke}, {Remillard}  \& {Veitch}}{{Blackburn}
  et~al.}{2015}]{Blackburn2015}
{Blackburn} L.,  {Briggs} M.~S.,  {Camp} J.,  {Christensen} N.,  {Connaughton}
  V.,  {Jenke} P.,  {Remillard} R.~A.,   {Veitch} J.,  2015, \mn@doi [\apjs]
  {10.1088/0067-0049/217/1/8}, \href
  {https://ui.adsabs.harvard.edu/abs/2015ApJS..217....8B} {217, 8}

\bibitem[\protect\citeauthoryear{{Cano}, {Wang}, {Dai}  \& {Wu}}{{Cano}
  et~al.}{2017}]{cano17}
{Cano} Z.,  {Wang} S.-Q.,  {Dai} Z.-G.,   {Wu} X.-F.,  2017, \mn@doi [Advances
  in Astronomy] {10.1155/2017/8929054}, \href
  {https://ui.adsabs.harvard.edu/abs/2017AdAst2017E...5C} {2017, 8929054}

\bibitem[\protect\citeauthoryear{{Cenko} et~al.,}{{Cenko}
  et~al.}{2012}]{cenko11}
{Cenko} S.~B.,  et~al., 2012, \mn@doi [\apj] {10.1088/0004-637X/753/1/77},
  \href {https://ui.adsabs.harvard.edu/abs/2012ApJ...753...77C} {753, 77}

\bibitem[\protect\citeauthoryear{{Cenko} et~al.,}{{Cenko}
  et~al.}{2015}]{cenko15}
{Cenko} S.~B.,  et~al., 2015, \mn@doi [\apjl] {10.1088/2041-8205/803/2/L24},
  \href {https://ui.adsabs.harvard.edu/abs/2015ApJ...803L..24C} {803, L24}

\bibitem[\protect\citeauthoryear{{Chevalier} \& {Li}}{{Chevalier} \&
  {Li}}{2000}]{Chevalier2000}
{Chevalier} R.~A.,  {Li} Z.-Y.,  2000, \mn@doi [\apj] {10.1086/308914}, \href
  {https://ui.adsabs.harvard.edu/abs/2000ApJ...536..195C} {536, 195}

\bibitem[\protect\citeauthoryear{{Daigne} \& {Mochkovitch}}{{Daigne} \&
  {Mochkovitch}}{1998}]{Daigne1998}
{Daigne} F.,  {Mochkovitch} R.,  1998, \mn@doi [\mnras]
  {10.1046/j.1365-8711.1998.01305.x}, \href
  {https://ui.adsabs.harvard.edu/abs/1998MNRAS.296..275D} {296, 275}

\bibitem[\protect\citeauthoryear{{Fong}, {Berger}, {Margutti}  \&
  {Zauderer}}{{Fong} et~al.}{2015}]{fong15}
{Fong} W.,  {Berger} E.,  {Margutti} R.,   {Zauderer} B.~A.,  2015, \mn@doi
  [\apj] {10.1088/0004-637X/815/2/102}, \href
  {https://ui.adsabs.harvard.edu/abs/2015ApJ...815..102F} {815, 102}

\bibitem[\protect\citeauthoryear{{Ghirlanda}, {Ghisellini}  \&
  {Nava}}{{Ghirlanda} et~al.}{2010}]{ghirlanda10}
{Ghirlanda} G.,  {Ghisellini} G.,   {Nava} L.,  2010, \mn@doi [\aap]
  {10.1051/0004-6361/200913980}, \href
  {https://ui.adsabs.harvard.edu/abs/2010A&A...510L...7G} {510, L7}

\bibitem[\protect\citeauthoryear{{Ghirlanda} et~al.,}{{Ghirlanda}
  et~al.}{2018}]{ghirlanda18}
{Ghirlanda} G.,  et~al., 2018, \mn@doi [\aap] {10.1051/0004-6361/201731598},
  \href {https://ui.adsabs.harvard.edu/abs/2018A&A...609A.112G} {609, A112}

\bibitem[\protect\citeauthoryear{{Ghirlanda}, {Salafia}, {Paragi}, {Giroletti}
  \& et al.}{{Ghirlanda} et~al.}{2019}]{Ghirlanda2019}
{Ghirlanda} G.,  {Salafia} O.~S.,  {Paragi} Z.,  {Giroletti} M.,   et al. 2019,
  \mn@doi [Science] {10.1126/science.aau8815}, \href
  {https://ui.adsabs.harvard.edu/abs/2019Sci...363..968G} {363, 968}

\bibitem[\protect\citeauthoryear{{Gill} \& {Granot}}{{Gill} \&
  {Granot}}{2018}]{Gill2018}
{Gill} R.,  {Granot} J.,  2018, \mn@doi [\mnras] {10.1093/mnras/sty1214}, \href
  {https://ui.adsabs.harvard.edu/abs/2018MNRAS.478.4128G} {478, 4128}

\bibitem[\protect\citeauthoryear{{Goldstein}, {Burns}, {Hamburg},
  {Connaughton}, {Veres}, {Briggs}, {Hui}  \& {The GBM-LIGO
  Collaboration}}{{Goldstein} et~al.}{2016}]{Goldstein2016}
{Goldstein} A.,  {Burns} E.,  {Hamburg} R.,  {Connaughton} V.,  {Veres} P.,
  {Briggs} M.~S.,  {Hui} C.~M.,   {The GBM-LIGO Collaboration} 2016, arXiv
  e-prints, \href {https://ui.adsabs.harvard.edu/abs/2016arXiv161202395G} {p.
  arXiv:1612.02395}

\bibitem[\protect\citeauthoryear{{Goldstein} et~al.,}{{Goldstein}
  et~al.}{2019}]{goldstein19}
{Goldstein} A.,  et~al., 2019, arXiv e-prints, \href
  {https://ui.adsabs.harvard.edu/abs/2019arXiv190312597G} {p. arXiv:1903.12597}

\bibitem[\protect\citeauthoryear{{Gottlieb}, {Nakar}, {Piran}  \&
  {Hotokezaka}}{{Gottlieb} et~al.}{2018}]{gottleib18}
{Gottlieb} O.,  {Nakar} E.,  {Piran} T.,   {Hotokezaka} K.,  2018, \mn@doi
  [\mnras] {10.1093/mnras/sty1462}, \href
  {https://ui.adsabs.harvard.edu/abs/2018MNRAS.479..588G} {479, 588}

\bibitem[\protect\citeauthoryear{{Granot}, {Panaitescu}, {Kumar}  \&
  {Woosley}}{{Granot} et~al.}{2002}]{granot02}
{Granot} J.,  {Panaitescu} A.,  {Kumar} P.,   {Woosley} S.~E.,  2002, \mn@doi
  [\apjl] {10.1086/340991}, \href
  {https://ui.adsabs.harvard.edu/abs/2002ApJ...570L..61G} {570, L61}

\bibitem[\protect\citeauthoryear{{G{\"u}ver} \& {{\"O}zel}}{{G{\"u}ver} \&
  {{\"O}zel}}{2009}]{Guver2009}
{G{\"u}ver} T.,  {{\"O}zel} F.,  2009, \mn@doi [\mnras]
  {10.1111/j.1365-2966.2009.15598.x}, \href
  {https://ui.adsabs.harvard.edu/abs/2009MNRAS.400.2050G} {400, 2050}

\bibitem[\protect\citeauthoryear{{Ho} \& {Zwicky Transient Facility
  Collaboration}}{{Ho} \& {Zwicky Transient Facility
  Collaboration}}{2021}]{at2021any_2}
{Ho} A.~Y.~Q.,  {Zwicky Transient Facility Collaboration} 2021, GRB Coordinates
  Network, \href {https://ui.adsabs.harvard.edu/abs/2021GCN.29313....1H}
  {29313, 1}

\bibitem[\protect\citeauthoryear{{Ho} et~al.,}{{Ho} et~al.}{2020}]{ho20}
{Ho} A. Y.~Q.,  et~al., 2020, arXiv e-prints, \href
  {https://ui.adsabs.harvard.edu/abs/2020arXiv200610761H} {p. arXiv:2006.10761}

\bibitem[\protect\citeauthoryear{{Ho}, {Perley}, {Yao}  \& {Andreoni}}{{Ho}
  et~al.}{2021}]{at2021any}
{Ho} A.~Y.~Q.,  {Perley} D.~A.,  {Yao} Y.,   {Andreoni} I.,  2021, Transient
  Name Server AstroNote, \href
  {https://ui.adsabs.harvard.edu/abs/2021TNSAN..20....1H} {20, 1}

\bibitem[\protect\citeauthoryear{{Huang}, {Dai}  \& {Lu}}{{Huang}
  et~al.}{2002}]{huang02}
{Huang} Y.~F.,  {Dai} Z.~G.,   {Lu} T.,  2002, \mn@doi [\mnras]
  {10.1046/j.1365-8711.2002.05334.x}, \href
  {https://ui.adsabs.harvard.edu/abs/2002MNRAS.332..735H} {332, 735}

\bibitem[\protect\citeauthoryear{{Ivezi{\'c}} et~al.}{{Ivezi{\'c}}
  et~al.}{2019}]{lsst}
{Ivezi{\'c}} {\v{Z}}.,  et~al., 2019, \mn@doi [\apj]
  {10.3847/1538-4357/ab042c}, \href
  {https://ui.adsabs.harvard.edu/abs/2019ApJ...873..111I} {873, 111}

\bibitem[\protect\citeauthoryear{{Kasliwal}, {Nakar}, {Singer}, {Kaplan}  \& et
  al.}{{Kasliwal} et~al.}{2017}]{Kasliwal2017}
{Kasliwal} M.~M.,  {Nakar} E.,  {Singer} L.~P.,  {Kaplan} D.~L.,   et al. 2017,
  \mn@doi [Science] {10.1126/science.aap9455}, \href
  {https://ui.adsabs.harvard.edu/abs/2017Sci...358.1559K} {358, 1559}

\bibitem[\protect\citeauthoryear{{Kathirgamaraju}, {Barniol Duran}  \&
  {Giannios}}{{Kathirgamaraju} et~al.}{2018}]{Kathirgamaraju2018}
{Kathirgamaraju} A.,  {Barniol Duran} R.,   {Giannios} D.,  2018, \mn@doi
  [\mnras] {10.1093/mnrasl/slx175}, \href
  {https://ui.adsabs.harvard.edu/abs/2018MNRAS.473L.121K} {473, L121}

\bibitem[\protect\citeauthoryear{{Kobayashi}, {Piran}  \& {Sari}}{{Kobayashi}
  et~al.}{1997}]{Kobayashi1997}
{Kobayashi} S.,  {Piran} T.,   {Sari} R.,  1997, \mn@doi [\apj]
  {10.1086/512791}, \href
  {https://ui.adsabs.harvard.edu/abs/1997ApJ...490...92K} {490, 92}

\bibitem[\protect\citeauthoryear{{Kumar}}{{Kumar}}{1999}]{kumar99}
{Kumar} P.,  1999, \mn@doi [\apjl] {10.1086/312265}, \href
  {https://ui.adsabs.harvard.edu/abs/1999ApJ...523L.113K} {523, L113}

\bibitem[\protect\citeauthoryear{{Lamb} \& {Kobayashi}}{{Lamb} \&
  {Kobayashi}}{2016}]{lamb_kobayashi16}
{Lamb} G.~P.,  {Kobayashi} S.,  2016, \mn@doi [\apj]
  {10.3847/0004-637X/829/2/112}, \href
  {https://ui.adsabs.harvard.edu/abs/2016ApJ...829..112L} {829, 112}

\bibitem[\protect\citeauthoryear{{Lamb}, {Mandel}  \& {Resmi}}{{Lamb}
  et~al.}{2018}]{lamb18_gw170817}
{Lamb} G.~P.,  {Mandel} I.,   {Resmi} L.,  2018, \mn@doi [\mnras]
  {10.1093/mnras/sty2196}, \href
  {https://ui.adsabs.harvard.edu/abs/2018MNRAS.481.2581L} {481, 2581}

\bibitem[\protect\citeauthoryear{{Lazzati}, {Morsony}, {Margutti}  \&
  {Begelman}}{{Lazzati} et~al.}{2013}]{lazzati13}
{Lazzati} D.,  {Morsony} B.~J.,  {Margutti} R.,   {Begelman} M.~C.,  2013,
  \mn@doi [\apj] {10.1088/0004-637X/765/2/103}, \href
  {https://ui.adsabs.harvard.edu/abs/2013ApJ...765..103L} {765, 103}

\bibitem[\protect\citeauthoryear{{Lazzati}, {Perna}, {Morsony}, {Lopez-Camara},
  {Cantiello}, {Ciolfi}, {Giacomazzo}  \& {Workman}}{{Lazzati}
  et~al.}{2018}]{Lazzati2018}
{Lazzati} D.,  {Perna} R.,  {Morsony} B.~J.,  {Lopez-Camara} D.,  {Cantiello}
  M.,  {Ciolfi} R.,  {Giacomazzo} B.,   {Workman} J.~C.,  2018, \mn@doi [\prl]
  {10.1103/PhysRevLett.120.241103}, \href
  {https://ui.adsabs.harvard.edu/abs/2018PhRvL.120x1103L} {120, 241103}

\bibitem[\protect\citeauthoryear{{Lee}, {Bartos}, {Privon}, {Rose}  \&
  {Torrey}}{{Lee} et~al.}{2020}]{lee20}
{Lee} K.~H.,  {Bartos} I.,  {Privon} G.~C.,  {Rose} J.~C.,   {Torrey} P.,
  2020, arXiv e-prints, \href
  {https://ui.adsabs.harvard.edu/abs/2020arXiv200700563L} {p. arXiv:2007.00563}

\bibitem[\protect\citeauthoryear{{Liang}, {Yi}, {Zhang}, {L{\"u}}, {Zhang}  \&
  {Zhang}}{{Liang} et~al.}{2010}]{liang10}
{Liang} E.-W.,  {Yi} S.-X.,  {Zhang} J.,  {L{\"u}} H.-J.,  {Zhang} B.-B.,
  {Zhang} B.,  2010, \mn@doi [\apj] {10.1088/0004-637X/725/2/2209}, \href
  {https://ui.adsabs.harvard.edu/abs/2010ApJ...725.2209L} {725, 2209}

\bibitem[\protect\citeauthoryear{{Lithwick} \& {Sari}}{{Lithwick} \&
  {Sari}}{2001}]{lithwick01}
{Lithwick} Y.,  {Sari} R.,  2001, \mn@doi [\apj] {10.1086/321455}, \href
  {https://ui.adsabs.harvard.edu/abs/2001ApJ...555..540L} {555, 540}

\bibitem[\protect\citeauthoryear{{Lloyd-Ronning} \& {Zhang}}{{Lloyd-Ronning} \&
  {Zhang}}{2004}]{lloyd04}
{Lloyd-Ronning} N.~M.,  {Zhang} B.,  2004, \mn@doi [\apj] {10.1086/423026},
  \href {https://ui.adsabs.harvard.edu/abs/2004ApJ...613..477L} {613, 477}

\bibitem[\protect\citeauthoryear{{Mandel}, {Farr}  \& {Gair}}{{Mandel}
  et~al.}{2019}]{mandel19}
{Mandel} I.,  {Farr} W.~M.,   {Gair} J.~R.,  2019, \mn@doi [\mnras]
  {10.1093/mnras/stz896}, \href
  {https://ui.adsabs.harvard.edu/abs/2019MNRAS.486.1086M} {486, 1086}

\bibitem[\protect\citeauthoryear{{Marcote}, {Nimmo}, {Salafia}, {Paragi},
  {Hessels}, {Petroff}  \& {Karuppusamy}}{{Marcote} et~al.}{2019}]{marcote19}
{Marcote} B.,  {Nimmo} K.,  {Salafia} O.~S.,  {Paragi} Z.,  {Hessels} J.~W.~T.,
   {Petroff} E.,   {Karuppusamy} R.,  2019, \mn@doi [\apjl]
  {10.3847/2041-8213/ab1aad}, \href
  {https://ui.adsabs.harvard.edu/abs/2019ApJ...876L..14M} {876, L14}

\bibitem[\protect\citeauthoryear{{Matsumoto}, {Nakar}  \& {Piran}}{{Matsumoto}
  et~al.}{2019}]{matsumoto19}
{Matsumoto} T.,  {Nakar} E.,   {Piran} T.,  2019, \mn@doi [\mnras]
  {10.1093/mnras/sty3200}, \href
  {https://ui.adsabs.harvard.edu/abs/2019MNRAS.483.1247M} {483, 1247}

\bibitem[\protect\citeauthoryear{{Meegan} et~al.,}{{Meegan}
  et~al.}{2009}]{fermi09}
{Meegan} C.,  et~al., 2009, \mn@doi [\apj] {10.1088/0004-637X/702/1/791}, \href
  {https://ui.adsabs.harvard.edu/abs/2009ApJ...702..791M} {702, 791}

\bibitem[\protect\citeauthoryear{{Mooley} et~al.,}{{Mooley}
  et~al.}{2018}]{mooley18_superluminal}
{Mooley} K.~P.,  et~al., 2018, \mn@doi [\nat] {10.1038/s41586-018-0486-3},
  \href {https://ui.adsabs.harvard.edu/\#abs/2018Natur.561..355M} {561, 355}

\bibitem[\protect\citeauthoryear{{Nakar}}{{Nakar}}{2007}]{nakar07}
{Nakar} E.,  2007, \mn@doi [\physrep] {10.1016/j.physrep.2007.02.005}, \href
  {https://ui.adsabs.harvard.edu/abs/2007PhR...442..166N} {442, 166}

\bibitem[\protect\citeauthoryear{{Nakar}}{{Nakar}}{2019}]{nakar19}
{Nakar} E.,  2019, arXiv e-prints, \href
  {https://ui.adsabs.harvard.edu/abs/2019arXiv191205659N} {p. arXiv:1912.05659}

\bibitem[\protect\citeauthoryear{{Nakar} \& {Piran}}{{Nakar} \&
  {Piran}}{2017}]{nakar17_cocoon}
{Nakar} E.,  {Piran} T.,  2017, \mn@doi [\apj] {10.3847/1538-4357/834/1/28},
  \href {https://ui.adsabs.harvard.edu/abs/2017ApJ...834...28N} {834, 28}

\bibitem[\protect\citeauthoryear{{Nakar}, {Piran}  \& {Granot}}{{Nakar}
  et~al.}{2002}]{Nakar2002}
{Nakar} E.,  {Piran} T.,   {Granot} J.,  2002, \mn@doi [\apj] {10.1086/342791},
  \href {https://ui.adsabs.harvard.edu/abs/2002ApJ...579..699N} {579, 699}

\bibitem[\protect\citeauthoryear{{Nakar}, {Ando}  \& {Sari}}{{Nakar}
  et~al.}{2009}]{Nakar2009}
{Nakar} E.,  {Ando} S.,   {Sari} R.,  2009, \mn@doi [\apj]
  {10.1088/0004-637X/703/1/675}, \href
  {https://ui.adsabs.harvard.edu/abs/2009ApJ...703..675N} {703, 675}

\bibitem[\protect\citeauthoryear{{Nava} et~al.,}{{Nava}
  et~al.}{2014}]{Nava2014}
{Nava} L.,  et~al., 2014, \mn@doi [\mnras] {10.1093/mnras/stu1451}, \href
  {https://ui.adsabs.harvard.edu/abs/2014MNRAS.443.3578N} {443, 3578}

\bibitem[\protect\citeauthoryear{{Racusin}, {Oates}, {Schady}, {Burrows}  \& et
  al.}{{Racusin} et~al.}{2011}]{Racusin2011}
{Racusin} J.~L.,  {Oates} S.~R.,  {Schady} P.,  {Burrows} D.~N.,   et al. 2011,
  \mn@doi [\apj] {10.1088/0004-637X/738/2/138}, \href
  {https://ui.adsabs.harvard.edu/abs/2011ApJ...738..138R} {738, 138}

\bibitem[\protect\citeauthoryear{{Rhoads}}{{Rhoads}}{1997}]{Rhoads1997}
{Rhoads} J.~E.,  1997, \mn@doi [\apjl] {10.1086/310876}, \href
  {https://ui.adsabs.harvard.edu/abs/1997ApJ...487L...1R} {487, L1}

\bibitem[\protect\citeauthoryear{{Rhoads}}{{Rhoads}}{2003}]{rhoads03}
{Rhoads} J.~E.,  2003, \mn@doi [\apj] {10.1086/368125}, \href
  {https://ui.adsabs.harvard.edu/abs/2003ApJ...591.1097R} {591, 1097}

\bibitem[\protect\citeauthoryear{{Ridnaia} et~al.,}{{Ridnaia}
  et~al.}{2020}]{konus_upperlim}
{Ridnaia} A.,  et~al., 2020, GRB Coordinates Network, \href
  {https://ui.adsabs.harvard.edu/abs/2020GCN.27039....1R} {27039, 1}

\bibitem[\protect\citeauthoryear{Ruderman}{Ruderman}{1975}]{ruderman75}
Ruderman M.,  1975, \mn@doi [Annals of the New York Academy of Sciences]
  {https://doi.org/10.1111/j.1749-6632.1975.tb31430.x}, 262, 164

\bibitem[\protect\citeauthoryear{{Ryan}, {van Eerten}, {Piro}  \&
  {Troja}}{{Ryan} et~al.}{2020}]{afterglowpy}
{Ryan} G.,  {van Eerten} H.,  {Piro} L.,   {Troja} E.,  2020, \mn@doi [\apj]
  {10.3847/1538-4357/ab93cf}, \href
  {https://ui.adsabs.harvard.edu/abs/2020ApJ...896..166R} {896, 166}

\bibitem[\protect\citeauthoryear{{Salafia}, {Ravasio}, {Yang}, {An}  \& et
  al.}{{Salafia} et~al.}{2021}]{Salafia2021}
{Salafia} O.~S.,  {Ravasio} M.~E.,  {Yang} J.,  {An} T.,   et al. 2021, arXiv
  e-prints, \href {https://ui.adsabs.harvard.edu/abs/2021arXiv210607169S} {p.
  arXiv:2106.07169}

\bibitem[\protect\citeauthoryear{{Sari} \& {Piran}}{{Sari} \&
  {Piran}}{1999}]{saripiran99}
{Sari} R.,  {Piran} T.,  1999, \mn@doi [\apj] {10.1086/307508}, \href
  {https://ui.adsabs.harvard.edu/abs/1999ApJ...520..641S} {520, 641}

\bibitem[\protect\citeauthoryear{{Singer}, {Ahumada}, {Ho}, {Zwicky Transient
  Facility (ZTF)}  \& {Global Relay Of Observatories Watching Transients Happen
  (Growth) Collaboration}}{{Singer} et~al.}{2020}]{singer20}
{Singer} L.~P.,  {Ahumada} T.,  {Ho} A. Y.~Q.,  {Zwicky Transient Facility
  (ZTF)}  {Global Relay Of Observatories Watching Transients Happen (Growth)
  Collaboration} 2020, GRB Coordinates Network, \href
  {https://ui.adsabs.harvard.edu/abs/2020GCN.26968....1S} {26968, 1}

\bibitem[\protect\citeauthoryear{{Speagle}}{{Speagle}}{2020}]{dynesty}
{Speagle} J.~S.,  2020, \mn@doi [\mnras] {10.1093/mnras/staa278}, \href
  {https://ui.adsabs.harvard.edu/abs/2020MNRAS.493.3132S} {493, 3132}

\bibitem[\protect\citeauthoryear{{Troja} et~al.,}{{Troja}
  et~al.}{2017}]{troja17_xrays}
{Troja} E.,  et~al., 2017, \mn@doi [\nat] {10.1038/nature24290}, \href
  {https://ui.adsabs.harvard.edu/\#abs/2017Natur.551...71T} {551, 71}

\bibitem[\protect\citeauthoryear{{Troja} et~al.,}{{Troja}
  et~al.}{2018}]{Troja2018}
{Troja} E.,  et~al., 2018, \mn@doi [\mnras] {10.1093/mnrasl/sly061}, \href
  {https://ui.adsabs.harvard.edu/abs/2018MNRAS.478L..18T} {478, L18}

\bibitem[\protect\citeauthoryear{{Wang} et~al.,}{{Wang} et~al.}{2015}]{wang15}
{Wang} X.-G.,  et~al., 2015, \mn@doi [\apjs] {10.1088/0067-0049/219/1/9}, \href
  {https://ui.adsabs.harvard.edu/abs/2015ApJS..219....9W} {219, 9}

\bibitem[\protect\citeauthoryear{{Zhang}}{{Zhang}}{2018}]{zhang_book}
{Zhang} B.,  2018, {The Physics of Gamma-Ray Bursts},
  \mn@doi{10.1017/9781139226530.
}

\bibitem[\protect\citeauthoryear{{Zhang} \& {Yan}}{{Zhang} \&
  {Yan}}{2011}]{zhang11}
{Zhang} B.,  {Yan} H.,  2011, \mn@doi [\apj] {10.1088/0004-637X/726/2/90},
  \href {https://ui.adsabs.harvard.edu/abs/2011ApJ...726...90Z} {726, 90}

\bibitem[\protect\citeauthoryear{{Zhang} et~al.,}{{Zhang}
  et~al.}{2007}]{zhang07}
{Zhang} B.,  et~al., 2007, \mn@doi [\apj] {10.1086/510110}, \href
  {https://ui.adsabs.harvard.edu/abs/2007ApJ...655..989Z} {655, 989}

\makeatother
\end{thebibliography}
%%%%%%%%%%%%%%%%%%%%%%%%%%%%%%%%%%%%%%%%%%%%%%%%%%

%%%%%%%%%%%%%%%%% APPENDICES %%%%%%%%%%%%%%%%%%%%%

%\appendix


%%%%%%%%%%%%%%%%%%%%%%%%%%%%%%%%%%%%%%%%%%%%%%%%%%
\appendix
\section{off-axis afterglow}\label{app:offaxis}
\begin{figure}
    \includegraphics[width=0.5\textwidth]{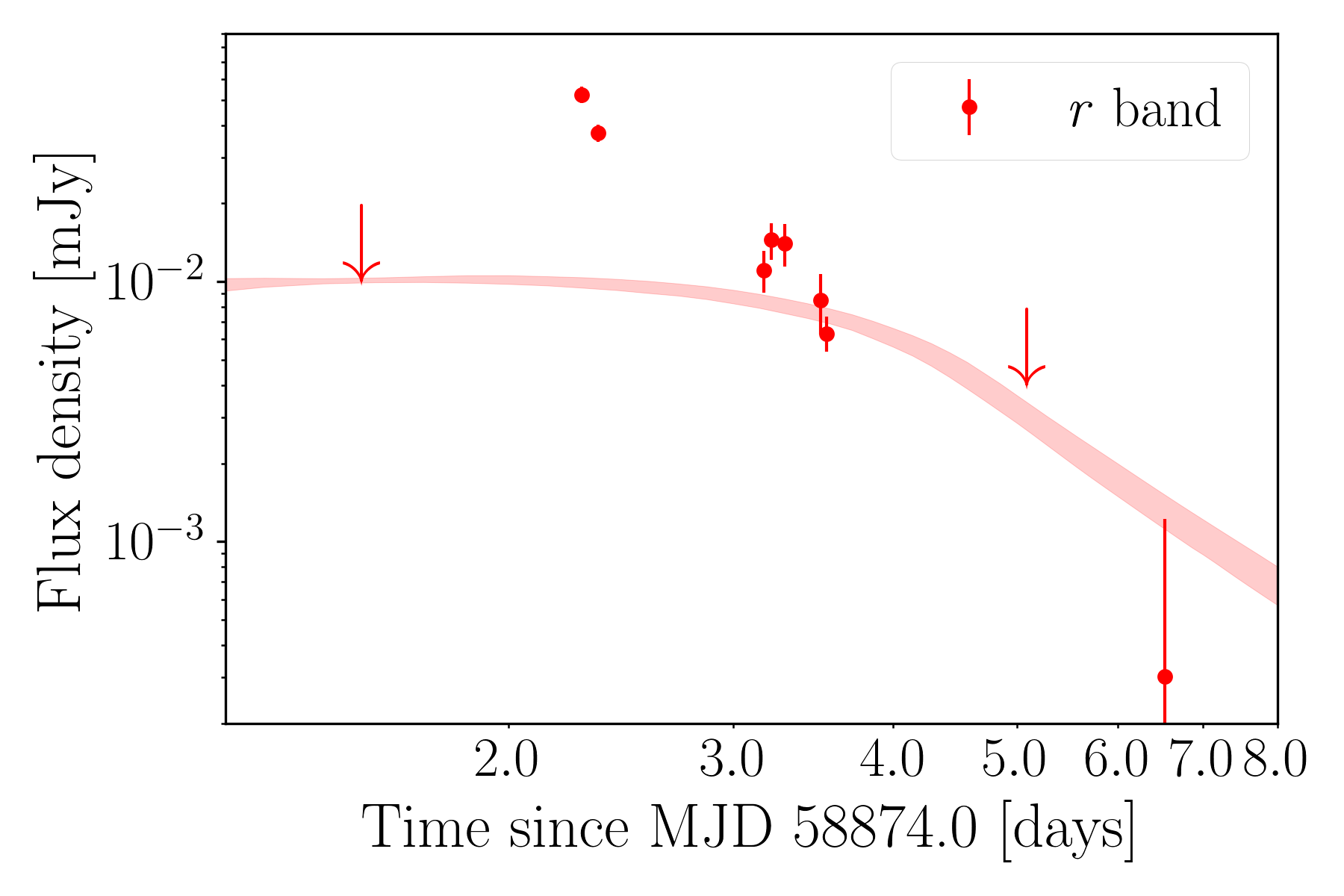} \hspace{-.4cm}
\caption{Data for AT2020blt in $r$ band, with arrows indicating upper limits. The red shaded curve is the $95\%$ credible interval predicted lightcurves for an off-axis structured jet model. The errors on the data are the measurement uncertainty reported by~\citet{ho20} combined with our modelled uncertainty for each model. We note that MJD 58874 is arbitrarily chosen to show the lightcurve more clearly. We also note that only $r$ band data is shown here but the fits are performed on the full multi-wavelength data.} 
\label{fig:offaxis}
\end{figure}

The lack of observed gamma-ray emission in AT2020blt could be a consequence of relativistic beaming, i.e., that we observed AT2020blt from outside the ultra-relativistic core.
We test this hypothesis by fitting the structured jet models above with a broad uniform prior on $t_{0}$ from 1 January 2019 to January 27.54 2020 (i.e., the first non-detection) and enforcing that the observer is located off-axis  i.e., $\theta_{\rm{observer}} > \theta_{\rm{core}}$. 
Our fit to the multi-wavelength data with the off-axis hypothesis is shown in Fig~\ref{fig:offaxis}.
As we discuss in Sec.~\ref{sec:afterglow}, an off-axis observer is unable to simultaneously explain the non-detection on January 27th, the rapid rise and subsequent rapid decay for all values of $\beta$ from $0.5-10$. We note that it is possible to explain the rapid rise and peak flux (but not the subsequent decay) for an observer located very close to the jet edge and with a jet to have $E_{k, \rm{iso}} \gtrsim \unit[10^{54}]{erg}$ which is uncomfortably large.
More quantitatively, we calculate a Bayes factor between a fit where the observer is forced off-axis and one where they are not. We calculate a Bayes factor of $\bfsploffaxis$ in favour of the on-axis hypothesis. The fit and the Bayes factor clearly indicate that the off-axis hypothesis is not able to explain the observations.

\section{Supplementary information}\label{app:supplementary}
\begin{figure*}
    \centering
    \begin{subfigure}
        \centering
        \includegraphics[width=85mm]{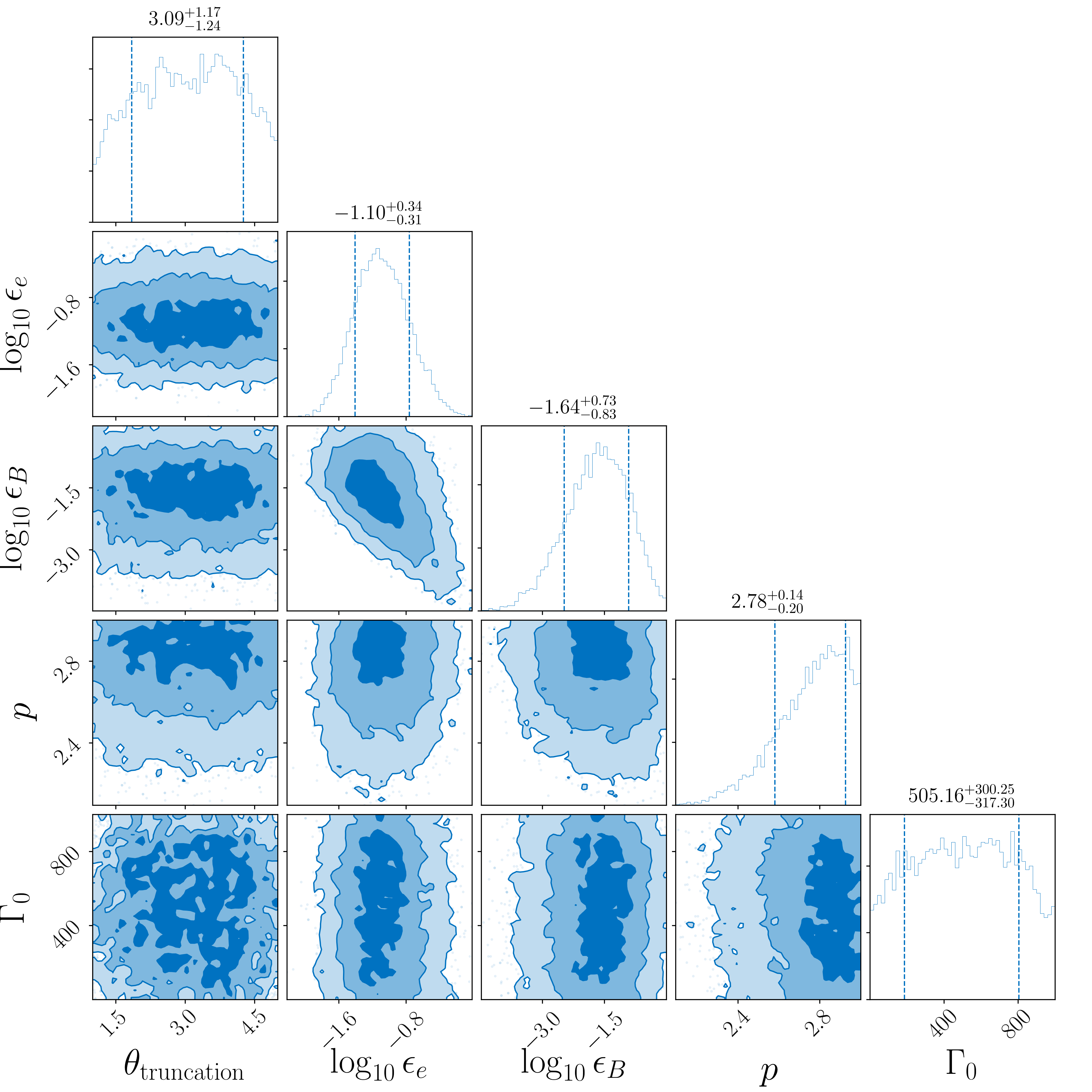} 
    \end{subfigure}
    \begin{subfigure}
        \centering
        \includegraphics[width=85mm]{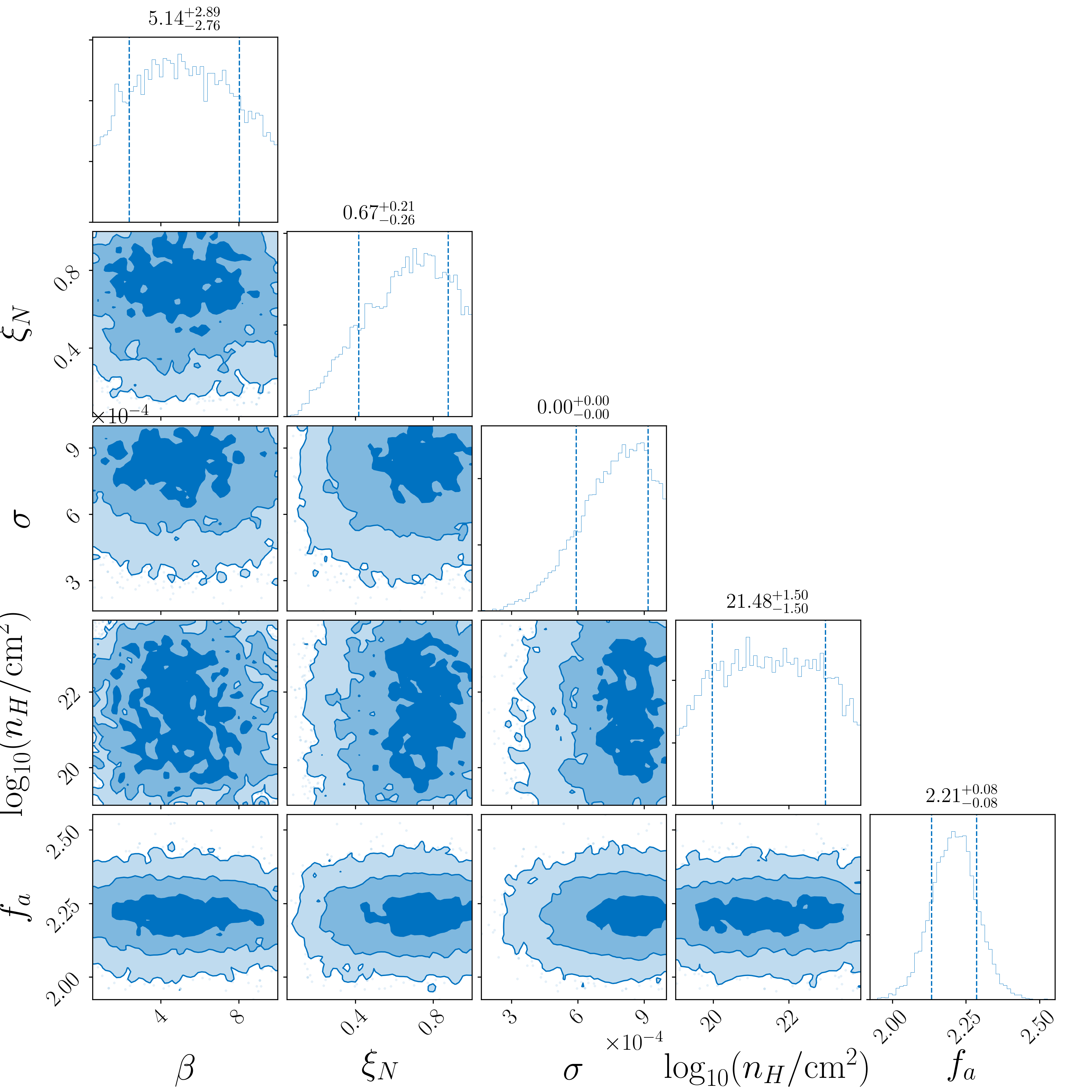} 
    \end{subfigure}
 \caption{One and two-dimensional posterior distribution for the structured jet model. The shaded contours indicate the $1-3 \sigma$ credible intervals. Note that the parameters not shown here are shown in Fig.~\ref{fig:t0corner} and~\ref{fig:corner}.}
    \label{fig:allcorners}
\end{figure*}

In Table~\ref{table:priors} we list the priors used in our analysis for the most preferred model i.e., the power-law structured jet along with a brief description of the parameters. In Fig.~\ref{fig:allcorners} we show the one and two-dimensional posterior distribution from our fit to the multi-wavelength data using the power-law structured jet model. We note that the parameters not shown here are previously shown in Fig.~\ref{fig:t0corner} and~\ref{fig:corner}.
\begin{table*}
\centering
 \caption{Parameters associated with the smooth power-law structured jet model along with a brief description and the prior used in our analysis. We note that the last two parameters $f_{a}$ and $\log_{10} (n_{H}/\unit{cm^{-2}})$ are used to account for the host galaxy extinction and are motivated by the analysis from~\citet{Guver2009}.}
 \label{table:priors}
 \begin{tabular}{lcc}
  \hline
  Parameter [unit] & Description & Prior \\
  \hline
$t_{0}$ [MJD] & burst time & $ \textrm{Uniform}[58872,58876.28]$\\
$t_{p}$ [MJD] & peak time i.e., the deceleration time & $ \textrm{Uniform}[58875.54,58876.28]$\\
$m$ & gradient of deceleration power law& $ \textrm{Uniform}[0.5,7]$\\
$\Gamma_0$ & initial Lorentz factor & $\textrm{Uniform}[1,1000]$ \\
$\theta_{\textrm{observer}}$ [rad] & observers viewing angle & $\textrm{Cosine}[0,0.7]$ \\
$\log_{10} (E_{\textrm{iso}}/\rm{erg}) $ & isotropic-equivalent energy & $\textrm{Uniform}[46,54]$\\
$\theta_{\textrm{core}}$ [rad] & half-width of jet core& $\textrm{Uniform}[0.02,0.2]$ \\
$\theta_{\textrm{wing}}$ & wing truncation angle of the jet &$\textrm{Uniform}[1,5] \times \theta_{\rm{core}}$\\
$\beta$ & power for power-law structure & $\textrm{Uniform}[0.5,10]$\\
$\log_{10} (n_{\textrm{ism}}/\unit{cm^{-3}})$ & number density of ISM & $\textrm{Uniform}[-3,4]$\\
$p$ & electron distribution power-law index & $\textrm{Uniform}[2,3]$\\
$\log_{10}\epsilon_{e}$ & thermal energy fraction in electrons& $\textrm{Uniform}[-5,0]$\\
$\log_{10}\epsilon_{b}$ & thermal energy fraction in magnetic field & $\textrm{Uniform}[-5,0]$\\
$\xi_{N}$& fraction of accelerated electrons & $\textrm{Uniform}[0,1]$\\
$\sigma$ & additional noise on measurements& $\textrm{Uniform}[10^{-4},10^{-3}]$\\
$\log_{10} (n_{H}/\unit{cm^{-2}})$ & hydrogen column density of the host galaxy & $\textrm{Uniform}[19,24]$\\
$f_{a}$ & extinction factor from~\citet{Guver2009} & $\textrm{Gaussian}(2.21,0.09)$\\
  \hline
 \end{tabular}
\end{table*}

% % Don't change these lines
\bsp    % typesetting comment
\label{lastpage}
\end{document}